\documentclass[sigconf]{acmart}

\usepackage{balance}
\usepackage{enumerate}
\usepackage{breakurl}

\copyrightyear{2021}
\acmYear{2021}
\setcopyright{acmcopyright}\acmConference[AIES '21]{Proceedings of the 2021 AAAI/ACM Conference on AI, Ethics, and Society}{May 19--21, 2021}{Virtual Event, USA}
\acmBooktitle{Proceedings of the 2021 AAAI/ACM Conference on AI, Ethics, and Society (AIES '21), May 19--21, 2021, Virtual Event, USA}
\acmPrice{15.00}
\acmDOI{10.1145/3461702.3462518}
\acmISBN{978-1-4503-8473-5/21/05}

\begin{document}
\fancyhead{}

\title{Reflexive Design for Fairness and Other Human Values in Formal Models}


\author{Benjamin Fish}
\affiliation{
\institution{Mila - Quebec AI Institute}
}
\email{benjamin.fish@mila.quebec}

\author{Luke Stark}
\affiliation{
\institution{University of Western Ontario}
}
\email{cstark23@uwo.ca}

%
%
%
%

\renewcommand{\shortauthors}{Fish and Stark}

\begin{abstract}

Algorithms and other formal models purportedly incorporating human values like fairness have grown increasingly popular in computer science. In response to sociotechnical challenges in the use of these models, designers and researchers have taken widely divergent positions on how formal models incorporating aspects of human values should be used: encouraging their use, moving away from them, or ignoring the normative consequences altogether.  In this paper, we seek to resolve these divergent positions by identifying the main conceptual limits of formal modeling, and develop four reflexive values--value fidelity, appropriate accuracy, value legibility, and value contestation--vital for incorporating human values adequately into formal models. We then provide a brief methodology for reflexively designing formal models incorporating human values.

\end{abstract}



\maketitle

\section{Introduction}

Formal modeling is a practice that underpins artificial intelligence (AI), and computer science more broadly. By formal modeling, we mean the application of mathematical structures, formulae, and algorithms to investigate, predict, or decide on a course of action. Computing deploys formal models liberally, but formal modeling is an area in which there has been relatively little attention to how model design is shaped by human norms, ethics, and values. Yet the problem of understanding how and why model design bears values is especially acute today, both because of the ubiquity of formal models as part of AI and machine learning (ML) systems and because of public and scholarly attention to the ethics and values of such systems.

In this paper, we develop a reflexive methodological approach to the design of formal models, and make concrete design suggestions for formal modeling with values in mind that take into account the sociotechnical implications and challenges of such work. Scholarship exploring how digital technologies are shaped by particular human values has long existed at the margins of computer science \cite{Agre:1997ts,Friedman:1996tm,Suchman:2006uq}, but today, normative concepts like fairness, accountability and transparency are no longer neglected in ML and other areas of computer science. The social impacts of automated unfairness and other forms of discrimination in AI systems are of increasingly urgent public concern \cite{Eubanks:2018wv,Hutchinson:2019ce,Lee:2018ik,Woodruff:2018if}. Yet despite these developments, scholars, designers, and software engineers remain confronted with a thorny challenge: how to approach the development of technologies like machine learning in practice in ways that account for social norms, ethics, and values. In the words of one set of design theorists, there is ``substantial agreement that design is value-laden,'' but ``significant variation arises in understanding how and why design bears values'' \cite{JafariNaimi:2015en}. In machine learning, as in other domains of computing, the work of understanding how to model with human values in mind is easier said than done. 

Machine learning and other related researchers and practitioners have created formal models for human values like fairness, privacy, accountability, and other values. Unfortunately, the impacts of these ``fair'' computational models often fail to satisfy even their own limited criteria for fairness when deployed in everyday settings \cite{Eubanks:2018wv,Hoffmann:2019hr}. The design and use of these models has thus been sharply critiqued by a range of scholars in critical HCI, science and technology studies, computer science, and related fields \cite{Selbst:2019wf,Benthall:2019dp,JafariNaimi:2015en}. These criticisms range from noting the difficulty of using quantifiable metrics to support meaningful social change \cite{Keyes:2019vi} to calls for moving away from formal modeling in certain sectors altogether \cite{Stark:2019fs}. 

What guidance for the incorporation of human values into formal models can technologists, researchers, engineers, and others working on the creation and implementation of formal models in machine learning systems, and elsewhere, look for amidst these many pitfalls? The multiple trajectories around formal modeling are incompatible, and that incompatibility is a major problem for the algorithmic fairness community. Specific, reproducible methods for designing fair models are practically nonexistent, and modeling work in other areas of machine learning which focus on traditional computational goals like speed and accuracy often ignores its broader normative consequences \cite{Wang:2018ht}. 

In this paper we argue for an attention to what we call \emph{reflexive values} and reflexive design more broadly as fruitful approaches to navigate these challenges. Reflexive design practice \cite{Pihkala:2016ca}, which underpins fields such as participatory design, involves a researcher or designer's constant engagement with their own role in the process of creation, and their interdependent relationship with other actors. Reflexivity is thus ``an orientation engaged in practice and in its relations that cut[s] across design'' \cite{Pihkala:2016ca} (21) -- attention is focused on the process of design and the judgements entailed by it rather than solely on the object being designed.

Here, we draw on the literature on reflexive design practice \cite{Sengers:2005uk,Friedland:2011bm} and on existing strategies for designing technical systems with human values in mind \cite{Flanagan:2008wx,Friedman:2006vb,Flanagan:2014vd,Selbst:2019wf,Friedman:2019vu} to provide a set of practical design strategies for grappling with, and partially resolving, the many challenges around formally modeling human values.  While reflexivity is central by implication to much existing scholarship in algorithmic fairness, there is little prior scholarly work providing proactive guidance: that which exists (e.g. \cite{Selbst:2019wf}) has focused largely on issues that arise with a na\"ive approach to the design of formal models. Formal modeling is not always an appropriate approach to a particular problem \cite{Malik:2020ws}, and value definitions will always be contested among the various groups affecting and affected by these systems \cite{Flanagan:2014vd}. However, we argue in this paper that reflexive design principles and the reflexive values we draw from them point towards concrete ways to design and use formal models appropriately, in the service of understanding and supporting particular human values in particular computational milieus.

First, we synthesize the main arguments for and against developing formal models with human values in mind in order to highlight the limitations to modeling work, and draw on these conclusions about the limitations of formal modeling to propose a set of \emph{reflexive values} that modelers should use as goals -- ``meta-values'' -- to ensure their particular values of interest are reflected in the design of the model itself and its impacts when deployed.  Then we apply these reflexive values as basis for a concrete methodology for operationalizing abstract human values in formal models. 

Note that ``formal model,'' ``design,'' and ``human values" are all contested concepts, with variation in precisely how they are defined, but in this paper we do not make any claims that too heavily rely on one particular conceptualization over another. Reflexive design values and practices can be applied at all points in the lifecycle of the project, including by researchers, designers, engineers, and other practitioners. Incorporating reflexive design practice into the process of formal modeling benefits not only the development of these models in machine learning, but in any field where modeling is a method. Given that few methods for formal modeling with human values in mind exist, this methodology has the potential to improve normative outcomes across modeling's many areas of application.   

\section{Previous Approaches to Human Values in Formal Modeling}\label{sec:values}

Formal modeling is the application of mathematical structures and formulae to describe, investigate, predict, or decide; it is widely used to create and understand computational systems. Current machine learning approaches employ formal modeling in a variety of ways. Often, formal models represent both input data and the desired output.  Typically, machine learning practitioners use a generic, domain-independent model, e.g.\ binary classification or multi-armed bandits \cite{lattimore2020bandit}, chosen for its apparent similarity to the problem at hand.  This generic model will be outfitted with a specific notion of reward or metric of evaluation, such as accuracy, that attempts to capture the desired output.  Algorithms, novel or otherwise, are then tested and evaluated for their suitability to solve this specific task.  Of course, the process need not happen in this order, researchers or practitioners may first start with a framework like binary classification and see how it may apply to their goals, but either way, there will be a formalism that attempts to capture the problem at hand.

In practice, ``modeling'' may refer not only to mathematizations of a domain or task, but to the variety of goals and values a system designer may have: including not only accuracy but speed, privacy, fairness, etc.  Formal definitions of fairness and related values generally involve placing constraints on the classifier as part of the process of predicting a property, or class, of a data point \cite{Narayanan:2018uz,Verma:2018hw,CorbettDavies:2018ud}.  Modeling can be applied to a wide variety of natural and social phenomena, often extending from the task itself to pieces of the domain where the task is being conducted.  We can attempt to measure any of these goals as particular functions of the mathematical objects that make up the formal model of the task and domain; we refer to the latter as formal models of the values themselves. Since these distinct kinds of models are constructed using similar methodology, have similar goals, and are often conflated anyway in existing arguments for and against their use in the literature, here we refer to them all under the umbrella of formal models.  Fleischmann \cite{Fleischmann:2010tj, Fleischmann:2011ws, Fleischmann:2016ga} and co-authors have explored formal models as sites of human values, the values of modelers themselves, and the crosscutting institutional pressures which shape model development.  Philosophers of science and science and technology studies (STS) scholars have also long evinced a deep interest in the social values represented in and through formal models \cite{Fourcade:2016ge, MacKenzie:2016id}. 

Scholars exploring human values in scientific modeling often distinguish between epistemic values, or ``considerations that are relevant to assessing the extent to which a representation matches the world'' \cite{Elliott:2014uq} (1) and non-epistemic or contextual values ``reflect[ing] moral, societal or personal concerns in the application of scientific knowledge'' \cite{Mohamed:2020eo} (3). As Elliott and McKaughan argue, both types of values are invariably present in a model: ``It is a descriptive fact of ordinary scientific practice,'' they note, ``that models represent their targets with varying degrees of success and typically focus selectively on those factors that are necessary in order to achieve the purposes for which they are used'' \cite{Elliott:2014uq} (6). The work of feminist philosophers of science further underscores this observation \cite{Potochnik:2012jb}. 

However, in comparison to the large literature focused on critiques of formal modeling or understanding the normative impact of designers and design, relatively little work has focused on rethinking the technical development of formal models so that they are created with human values explicitly in mind. Scholarship seeking to improve design lifecycle in artificial intelligence has focused on changing incentives for participants: ensuring those impacted by the system are involved meaningfully in its creation at every stage \cite{Madaio:2020cn,Sloan:2020um}. 

Here we argue the process of formal modeling can itself be improved through reflexive design practices. It should not be controversial to state that models, including formal ones, are abstract representations, the development of which entail making choices shaped by the norms of scientific practice and of social life. Reflexivity entails ``the capacity of any entity to turn back on itself [and] to make itself its own object of investigation'' \cite{Pihkala:2016ca} (21), and is a hallmark of feminist STS scholarship, in particular to the notion of situated knowledges as instrumental to both scholarly and design work \cite{Haraway:1988us, Suchman:2002tp}. Reflexivity is also a key element of critical design practices such as participatory design \cite{Simonsen:2013ws}, as it involves ``an interest in and commitment to attending to intentions and interpretations, power and accountabilities'' \cite{Pihkala:2016ca} (22) in the course of an object's development. 

Work in the philosophy of technology and science and technology studies (STS), variously termed Values in Design (VID) \cite{Knobel:2011hza}, Value Sensitive Design \cite{Friedman:2006vb} or Values@Play \cite{Flanagan:2008wx,Flanagan:2014vd}, is similarly focused on the way design choices reflect and rebound on human values in technological development. As Flanagan and Nissenbaum \cite{Flanagan:2014vd} define them, values are ``properties of things and states of affairs that we care about and strive to attain'' (5). These methods consider the emergence of values from a sociotechnical perspective, attending both to human and non-human actors and to the technical, material, and social aspects of a given design context \cite{Shilton:2017hh}. VID as a scholarly tradition underpins much of the current literature in STS around algorithmic fairness \cite{Friedman:1996tm,Agre:1997tl,Anonymous:2016fr}, and VID principles have been applied to the design of various digital systems \cite{Stark:2013bl,Howe:2009uj,Shilton:2013dd}. This tradition is implicitly reflexive: as Shilton \cite{Shilton:2017hh} notes, researchers in the VID tradition, ``[increasingly] interpret values as processes or practices'' (249), which are emphasized or downplayed through material and technical specifications. Shilton \cite{Shilton:2018fb} describe VID work as inherently interdisciplinary, ``attaching the ongoing conversation about values, ethics, and politics of technologies concretely to design'' (119). Flanagan, Howe, and Nissenbaum observe that each element of a VID method is ``dynamically interconnected with the others and can affect and be affected by them in a reflexive manner'' \cite{Flanagan:2014vd} (340).

In response to the increased popularity of formal modeling as a solution to concerns around the ethics and fairness of machine learning and other AI systems, recent scholarship has drawn on the VID tradition to describe some of the methodological challenges in deploying such tools in the pursuit of algorithmic fairness \cite{Selbst:2019wf, Benthall:2019dp, Abebe:2018jc}. For instance, Selbst et al.\ \cite{Selbst:2019wf} develop existing work in STS to detail many of the challenges and pitfalls of formal modeling around values and provide a taxonomy of ``abstraction traps'': some of the many ways in which designers of algorithmic systems can fail to consider the broader sociotechnical context of their work in developing formal representations of the world. As those authors observe, ``To treat fairness and justice as terms that have meaningful application to technology separate from a social context is\ldots to make a category error'' \cite{Selbst:2019wf} (2). The authors suggest designers consider and remediate against each of these traps in turn to help ensure their systems do not become overly abstracted from the concerns of lived experience.

While salutary, much of this previous work fails to comprehensively do two things: a) answer the competing claims about the usefulness of formal models described above, and b) provide granular guidance for how incorporate reflexivity into the specifics of design practice. In the next sections, we engage with arguments from science and technology studies (STS), computer science, and related fields regarding the capacities and limits of formal modeling. Each provides a variety of perspectives on the degree to which human values may be usefully modeled, and by extension a variety of insights on how such modeling might be successful in practice. Out of this analysis, we suggest four reflexive meta-values---value fidelity, accuracy, value legibility, and value contestation---through which designers and developers can ground their modeling work. 

\section{Can human values be usefully modeled at all?}

The conceptual underpinnings of formal models contribute to their societal effects. More than simply a problem of biased input data or the individual values of systems designers, formal modeling shapes what social and ethical values such systems can support in practice. In the next sections, we interrogate the question of whether producing formal models involving human values has any worth at all, and whether this worth outweighs the intrinsic limits of these mechanisms in the first place. 

The production of mathematical definitions--that is, formal models--of human values like fairness, accountability, and transparency has been rife in recent work examining the social and ethical impacts of AI systems.  It is worth noting that categorizing and making comparisons are intrinsically a part of all systems of valuation: to value is to believe that some actions, processes, or simply ways of thinking are to be preferred over others, and some values will be foregrounded or placed first, while others are pushed to the background or placed last. But whether human values \emph{should} be modeled formally at all first depends on having clarity about what the formal model in question is supposed to accomplish. Such outcomes can be politically progressive ones. For instance, Abebe et al.\ \cite{Abebe:2020kr} argue there are examples of computational abstractions that can support social change, including clarifying through omission particular societal considerations otherwise ignored or misunderstood by policymakers. Formal models can represent relevant distinctions in different social domains schematically, and can help highlight legitimate disagreements we have in what abstract concepts like fairness actually mean.

If we accept that formal models of human values can have benefits, we must still examine the limits of formal models. In order for human values to be modeled appropriately, all those affected by the model should be able to draw meaningful conclusions about both the values at hand and the real world from the model. Elliott and McKaughan observe that ``[modeled] representations can be evaluated not only on the basis of the relations that they bear to the world but also in connection with the various uses to which they are put'' \cite{Elliott:2014uq} (3). In other words, an initial criterion for formally modeling values is whether the model in question both a) bears a reasonable relation to the human values it schematizes, and b) whether the model is used and useful for a purpose which in turn supports those same values. 

Knowing when these two criteria are fulfilled, however, is a challenge. For instance, JafariNaimi, Nathan, and Hargraves \cite{JafariNaimi:2015en} are critical of existing Values in Design methods, such as Value Sensitive Design \cite{Friedman:2006vb, Friedman:2019vu} and Values@Play \cite{Flanagan:2008wx, Flanagan:2014vd}, identifying what they describe as an ``Identify/Apply'' logic animating the process of translating values into designed objects. JafariNaimi et al.\ note that these methods ``share the implicit assumption that once the values have been accurately identified, they can be applied to the design of a technology that will in turn embody, bear, or advance those values'' (94). Critiquing this assumption as being removed from the practicalities of most design practice, JafariNaimi et al.\ argue that in the vast majority of VID methods, ``the work of understanding a value and the work of applying a value to design are separate'' (94), making such methods impractical for designers of any sort caught up in the situated experience of creating artifacts -- for instance, formal models -- from the ground up. 

JafariNaimi et al. argue instead for understanding values in a pluralist manner, as ``hypotheses by which to examine what the situation is, what the possible courses of action are, and how they might transform the situation'' in the context of design. In this view, values should not be applied, but instead ``serve situations as hypotheses [...] to bring forward the means that are available to advance valued ends and to advance understanding of the value of ends achievable through conceivable or available means'' (97). In a related vein, Seaver \cite{Seaver:2017if}  argues that the dominant view of formal models' effects on society, of an algorithm as discrete object affecting the world flowing around it akin to an algorithmic rock in a cultural river, is incorrect. Seaver notes this metaphor does not describe reality: in order to understand the effect of algorithms and other formal models, he suggests, they must be understood as ``composed of collective human practices'' \cite{Seaver:2017if} (5), and thus tied up with the granular particularities of everyday design practice.  Such reflexivity is implicitly a feature of many existing VID methods, particularly as applied in practice.  

We believe some formal models are better designed and deployed than others, meaning their design, deployment, and output can better reflect a value used as a hypothesis, but values cannot, as JafariNaimi et al. argue, ``be used as pre-established formulas that yield proper courses of action'' (97).  Yet is a modeler's subjective sense that their chosen design is related to the value on which they have drawn on as a hypothesis enough to say that it comports with that value? How should formal modelers in particular go about actually developing mathematical objects out of values-as-hypotheses? And what standard of evaluation should be used to determine if a model bears a reasonable relation to the human values it schematizes, and whether the model is useful for a purpose supporting those same values?

Here, we introduce several \emph{reflexive values} to provide guidance for the iterative evaluation of these hypotheses.  As we note in the introduction, reflexivity is ``an orientation engaged in practice and in its relations'' \cite{Pihkala:2016ca} (21). Reflexive design practice \cite{Sengers:2005uk,Friedland:2011bm} is a hallmark of participatory approaches to human-computer interaction work: Sengers et al. \cite{Sengers:2005uk} observe that ``reflexive recognition of the politics of design practice and a desire to speak to the needs of multiple constituencies in the design process'' are central to ensure shared attention is focused on the design process as a whole and on its impacts, rather than merely on the technical object being designed. It is important to observe, however, that reflexivity on its own is not a design panacea. That is why we suggest a number of what we term \emph{reflexive values} intended to help channel reflection into appropriate practice.  Formal modelers concerned with the normative impacts of their work thus need not face a double bind: while there is no ``pre-established formula'' for a value that can be concretized or formalized within a model, the normative impacts of particular models can be assessed by paying attention to the dynamics of these reflexive values. The first of these values is one we term ``value fidelity.''

\subsection{Value Fidelity}\label{sec:value_fidelity}

Following JafariNaimi et al. \cite{JafariNaimi:2015en}, we argue formal modeling that engages with human values is only appropriate when such models are developed in the context of particular, contingent applications. Restricting the use of particular formal models to particular social domains honors those domains' non-epistemic values \cite{Elliott:2014uq}, and ensures the model's approximate fidelity within the domain.  As Selbst et al.\ \cite{Selbst:2019wf} note, ``concepts such as `fairness' are not tied to specific objects but to specific social contexts,'' meaning that many formal models abstracting human values are not ``portable'' (10). Formal models should thus be assessed iteratively and reflexively as they are being developed in context for how particular values are being technically conceptualized and translated into the model, and evaluated in situ with regard to the values they purport to represent.  This process of reflexive assessment around how the technical mechanisms of a model represent the dynamics of desired values as they emerge in human practice we term \emph{value fidelity}.  

Paying close attention to the challenge of value fidelity as a reflexive design meta-value should help modelers navigate the fact that it is never immediately obvious how to treat an abstract value so it can be operationalized, or whether it should be operationalized in the first place. No model will in practice perfectly capture a real-world domain, and if a normative theory is also allusive, vague, or unclear, the challenge is compounded by the need for a second level of design clarification around what a particular value means in practice. This design process, exactly because it must enable particular choices for overcoming vagueness and abstraction of values, becomes crucial to forming particular domain specific conceptualizations of a value. The wide array of particular choices made in that process -- everything from how people are recruited into the design context to the particular mathematical structures employed in the model -- will invariably influence this conceptualization. A focus on value fidelity should also push modelers to explicitly articulate and concretize their own values in context, seeking to satisfy JafariNaimi et al.'s concern regarding the embedded nature of values in particular design areas \cite{JafariNaimi:2015en}.

Selbst et al.\ also observe the danger of labeling a particular algorithm, formal model or other technical element as ``fair'' regardless of its context \cite{Selbst:2019wf}. This point holds for values in general: a focus on value fidelity helps avoid the trap of assigning value to algorithms divorced from the sociotechnical situations enabling algorithms to have values in the first place \cite{Seaver:2017if}.   Given that, ``questions about how best to balance trade-offs between various desiderata clearly depend on what our goals are when we make our [modeling] choices'' \cite{Elliott:2014uq} (4), restricting the use of particular formal mechanisms and models to the particular social domains for which they were designed also makes sense in considering such models' epistemic values as well. Reflexivity across all these choices gives modelers a better chance of being ``in the ballpark.'' For instance, a modeler seeking to satisfy the conditions of Andersonian democratic equality \cite{Anderson:1999wh} (which are grounded in equal social standing) cannot, if thinking reflexively, simply apply a statistical measure of distributional group fairness and call it a day. Value fidelity is also a two-way street: some technologies are so conceptually harmful from the standpoint of human values that it makes little difference how sensitively they are designed or deployed \cite{Stark:2019fs}. 

%
%
%
%
%
%

 \subsection{Accuracy}\label{sec:accuracy}

Value fidelity emphasizes the need for modelers to restrict the use of particular formal models to the particular social domains for which they were designed, and to reflexively assess these restrictions and their impacts of modeling choices throughout the design process.   Fixing models to particular real-world domains in which designers must situate their values helps to focus modeling, pushing designers to weigh the valences of particular normative situations rather than leave it up to others when the model is deployed. Such real-world situations involve a) the many choices contingent on the design of the sociotechnical system, and b) what kind of decision-making body (judicial system, firm, organization within a firm, etc.) is responsible for upholding those decisions \cite{Nissenbaum:2011uv}.  Though a real-world domain comes equipped with this kind of information, what counts as an appropriate model for such a domain is not immediately obvious: after all, as the aphorism goes, all models are wrong but only some are useful \cite{Malik:2020ws}.

How to produce accurate models is a central question of virtually all scientific disciplines, and much of that literature is outside the scope of this work.  By assigning conceptual meaning to variables, a model asserts that the variable is an accurate depiction of that concept.  And if a model does not capture essential data about a real-world domain, any predictions or conclusions that can be drawn from it will inaccurate, and so any normative consequences it predicts will also not be right. Even ostensibly simple concepts in computer science like running time and space can be difficult to formalize, depending on the exact domain, and have long histories behind their modern conceptions \cite{FortnowH03}. And given that even well-conceptualized variables may be hard to observe, the use of proxy variables causes further problems.  How do we know a proxy variable is a good choice? And how do proxy variable choices violate desired values within a system? \cite{Jacobs:2019vx}. 

After ensuring fidelity to a particular social domain, designers should take active steps during the modeling process to delimit their models by actively constraining their parameters. These constraints should be in line with both the epistemic and non-epistemic values of the domain at hand, and involves a focus on our second reflexive meta-value, \emph{accuracy}. We believe it is necessary to scrutinize both the accuracy of the data used to build the model, and the mathematical structures and dynamics used as proxies for real-world processes. This process of reflexive, iterative assessment and evaluation should include, but is not limited to the accuracy of the output of a classifier or other predictor. It should also include the assumptions and claims the modeler has implicitly made in assigning semantics to mathematical objects. There are a variety of situational decisions in response to these questions, depending on the particular contexts of model design and deployment noted above. We suggest for one that normative values should guide modelers above excessively comprehensive accuracy; given that all models are partial, a simple model designed thoughtfully, reflexively, and explicably will likely support normative goals better than a more complex and kludgy one.

\section{What kind of claims can formal models make?}\label{sec:kinds_of_claims}

By virtue of the mathematical objects of which they are comprised, formal models make particular kinds of claims, both epistemological and non-epistemological. Ensuring both value fidelity and accuracy does not address the limits of these claims: to design successfully, we thus need to consider in detail what kinds of claims formal models make about the world.  Worse, as Selbst et al.\ \cite{Selbst:2019wf} put it, there is ``no way to arbitrate between irreconcilably conflicting definitions [of values] using purely mathematical means'' (5).

Green and Viljoen \cite{Green:2020boa} observe that in many formal models developed for AI systems, ``only considerations that are legible within the language of algorithms,'' such as efficiency and accuracy, ``are recognized as important design and evaluation considerations'' (23).  These ``internalist'' tendencies of machine learning models make algorithmic responses to questions concerning human values such as fairness problematic, because ``problems with quantification [affect] everything downstream'' \cite{Malik:2020ws} (3). Some concepts are easier to express in some languages than in others. For instance, Ananny \cite{Ananny:2015fn} argues that performing algorithmic categorization in social environments is particularly harmful because such categorizations constrain the available set of human actions; Hoffmann \cite{Hoffmann:2019hr} notes that in doing so, formal modeling reifies the categories used, even when those categories are grounded in harmful stereotypes.

Precisely because of their reliance on quantification, formal models, as Malik observes \cite{Malik:2020ws}, ``commit to working with proxies rather than the actual constructs of interest'' (2) (see also \cite{Jacobs:2019vx}).  Likewise, Selbst et  al.\ observe that formal models of values like fairness are invariably imperfect and simplified representations, unable to ``capture the full range of similar and overlapping notions of fairness and discrimination in philosophical, legal, and sociological contexts.'' Practitioners often deploy proxy variables, not only because of a lack of data, but also because some concepts are easier to express than others in formal language.  Proxies are used to avoid modeling of fuzzy social effects, to simplify by ignoring exogenous variables to the modeled system, to fix more easily manipulable proxies (such as maximizing profit) as stated human goals, or to simplify the number of different options available (such as discretizing input or output variables or assuming that social networks are composed of simple kinds of edges) \cite{Olteanu:2019el}.  

By holding fixed social and political factors through a particular choice of proxies, internalism narrows the scope of technological interventions, and risks maintaining the social and political status quo with significant normative consequences--including the potential to reinforce discrimination or other harmful social conditions. For instance, Eubanks points out that, ``when automated decision-making tools are not built to explicitly dismantle structural inequalities, their increased speed and vast scale intensify them dramatically'' \cite{Eubanks:2018uy, Eubanks:2018wv}. Green and Viljoen \cite{Green:2020boa} argue that even explicitly modeling social values will not solve this problem, because the model always has an internalist tendency. Selbst et al.\ also offer a series of observations intended to address the ``formalism'' trap, centered around the need to ``consider how different definitions of fairness, including mathematical formalisms, solve different groups' problems by addressing different contextual concerns'' (10). These authors are rightly concerned with the ways in which powerful social groups or vested interests can foreclose particular technical developments not because of their lack of fitness for purpose, but because, as the authors observe, ``the social world is a mechanism that fundamentally shapes technical development at every level.'' (11). 

Formal modeling as a way to understand the normative valences of a sociotechnical system is useless unless a model engages with externalities in the most comprehensive fashion possible. The tendency of automation to reify and extend existing power asymmetries is not captured by formal models, but is nonetheless central to their impact in the world. Green and Viljoen \cite{Green:2020boa} do suggest that expanding formal models to include social values, which they term ``formalist incorporation'', may still have occasional value: they observe such modeling may be situationally and strategically useful, while always noting such solutions are insufficient as full remedies to the inherent limitations of formal modeling. But if modelers go about formalist incorporation, how should they do it in the least problematic way possible? To navigate the pitfalls of formalist incorporation, we argue modelers should embrace the reflexive values of \emph{value legibility} and \emph{value contestation}.

\subsection{Value Legibility}\label{sec:value_legibility}

Most formal models prompt strong normative concerns whose effects are not observable within the model itself: they thus lack our third reflexive meta-value, what we term \emph{value legibility}. Often when scholars critique the gap between their conception of fairness and technical definitions of categories such as group fairness, what is implicitly at issue is such value legibility: the fact that effects or processes that are known to have normative impact are left out of a model's design, and/or are unaccounted for in its deployment. In effect, the negative normative consequences of these values are externalized, made illegible to formal methods but all too appreciable to those living with the real-world impacts of modeling. For instance, Hoffman \cite{Hoffmann:2019hr} notes even algorithmic fairness techniques for attempting to protect large classes of protected attributes, such as those proposed by Kearns et al. \cite{Kearns:2018wp}, ``fall short as an intersectional approach'' \cite{Hoffmann:2019hr}  (906). Hoffmann observes that, ``intersectionality is not a matter of randomly combining infinite variables to see what `disadvantages' fall out; rather, it is about mapping the production and contingency of social categories'' \cite{Hoffmann:2019hr} (906). Intersectionality, in other words, is a social concept unobservable within even some sophisticated formal models.  

Reflexive design practice helps to flag issues of value legibility: modelers must either formalize effects or processes that are understood to have normative impact in the model itself, or that when it is not practical or even possible to do, deal with these effects in another fashion rather than ignoring them.  Conclusions or decisions made by a model can only use the information inside of the model, which means any normative claims drawn from the model must also be a function solely of the information contained inside that model. While the problem of internalism forces our hand to try to make values observable via the model, the need for value legibility is also valuable in clarifying the practical impacts of particular models.  As Abebe et al.\ \cite{Abebe:2020kr} puts it, citing Kleinberg et al.\ \cite{kleinberg2016guide}, ``algorithms may help to lay bare the stakes of decision-making and may give people an opportunity to directly confront and contest the values these systems encode.''  Making the consequence of modeling choices as concrete as possible allows for the stakes of design decisions to be as clear as possible to all concerned, both in terms of the model itself and around its potential effects. 

\subsection{Value Contestation}\label{sec:contestability}

No formal model can guarantee the values its designers have and seek to embody in their designs will translate into real-world effects. However, the values of system designers do produce path dependencies \cite{Hughes:1987tv} that can constrain the range of value outcomes for a particular model. Formal models also concentrate power in the hands of those same designers, making it likely only those with the access and knowledge necessary to understand and create algorithmic approaches will be able to decide who and what is valued in the system. Power asymmetries are not solely an issue for algorithm design or computational systems; they pertain to any system that is part of a decision-making apparatus that affects other people. Approaches to this element of the problem must be subject to the same, old-fashioned political processes that would guide any kind of sociotechnical decision-making \cite{Sloan:2020um}. For instance, Selbst et al.\ call for attention to the social groups that both do and do not have a voice to influence the design process.  Yet what should all participants in the design of formal models actually do, and not do, once so involved? How should designers proceed as relevant voices are heard?  And what should modelers do if stakeholders disagree with them or each other? 

Models do not just encode assertions about reality (which in turn may have normative consequences). Models also encode the modelers' assertions and assumptions about normative judgements directly into the model. This problem is tied tightly to the issue of proxy measures, and how these measures become reified as ground truth:  for instance, in models which implicitly accept racial categories as fixed, inherent properties of a person \cite{Benthall:2019dp}.  Designers must reject these assumptions and embrace our fourth reflexive value, \emph{value contestation} in the design process. It is not enough to recognize that human values are contestable; modelers must be aware that they always take part in the contest. Given that even reasonable people will inevitably disagree about values and how to conceptualize them, there will always be sites of normative contestation -- what Katie Shilton terms ``values seams'' \cite{Shilton:2017hh}. 

Pihkala and Karasti \cite{Pihkala:2016ca} observe that ``being reflexively engaged posits the designer-researcher as involved, as a participant'' (28), suggesting the need for ``participation in plural'' -- forcing constant reconsideration not only of which values are animating a model, but also who is being heard and heeded as the model is being developed.  Value contestation is closely tied to contestable social concepts:  for instance, implicitly accepting other people's decisions regarding what counts as ``crime'' is a judgement call which is influenced by the record keepers' incentives, social structure, and other social factors. Value contestation is vital because the sites where it happens are particularly easy places to sneak unwanted assumptions into formal models.  In questioning what has been held fixed, often social and political factors, value contestation is also an antidote to unreflexive internalism.  There are many possible ways to actually perform such contestation: formalist incorporation -- appending or modifying the model itself -- is not necessarily inappropriate, as long as it deals with the normative judgements at stake and as long as such processes are always understood as contested.

\section{Reflexive Values in Model Design Practice}\label{sec:method}

Formal models cannot simply be infused with abstract human values, are dependent on proxies to make claims, and are most easily able to represent a narrow range of values. The reflexive meta-values we propose can help anyone working on developing formal models to cope with these conceptual limitations, holding valuable lessons for technologists and designers seeking to develop methods for formally modeling human values, and incorporating such models into practical applications when and where appropriate. Synthesizing the discussion above, we summarize out the following lessons:

\begin{enumerate}

\item As JafariNaimi et al. note, human values are often presented as abstract, but are always enacted through particular design contexts and should be treated as such at every stage of the design and evaluation process.  This process requires reflexive assessment of how the technical mechanisms of a model represent the dynamics of desired values as they emerge in human practice -- what we call \emph{value fidelity}. 

 \item Formal models are representations of reality, but are often insufficiently proximate to real-world conditions to draw reliable conclusions the best way to situate and operationalize human values in the design of the model--a problem of \emph{accuracy} both of proxies and how proxy data interacts with the mechanisms of the model itself.

\item Without a high degree of design reflexivity, the broader normative consequences of a formal model's design and deployment are often either not modeled, not considered, or not modelable at all--a question of what we term \emph{value legibility}.

\item Even with a high degree of design reflexivity, formal models are designed with subjective assumptions about values that can lead to undesirable normative consequences, or at the very least conflicts around the normative valence of particular models (what we term \emph{value contestation}).

\end{enumerate}

Below, we provide a basic methodology for formal model design grounded in these four reflexive meta-values. Modelers following this method should be more aware of a) how they are shaping the effects of the system, and b) how the deployed system will impact those using it than they would be using traditional agile development techniques or even older ``waterfall'' development processes \cite{Polonetsky:2017vl}. Inspired by existing VID methods \cite{Flanagan:2008wx,Flanagan:2014vd}, we have divided our method into a pre-design stage, a design stage, and a post-design stage. Perhaps the most critical element of pre-design is the need to assess whether it is appropriate to design or deploy a formal model in the first place \cite{Baumer:2011uj}.

\subsection{Pre-design}\label{sec:predesign}

Given that human values always enacted in particular design contexts and should be treated as such at every stage of the design and evaluation process, respecting value fidelity means modelers should begin their reflexive assessment of their work long before they reach for pen and paper or they start coding. This assessment should start with reflection on the makeup of the modeling team itself.  Attention to our reflexive meta-values in the context of agile development means the design process for creating a particular formal model is likely not possible in practice with one designer, or even several designers with a homogenous perspective. Having a variety of expertise involved is a consequence of the fact that it is almost always more appropriate to have the stakeholders in the system -- those who will be most affected by its deployment and use of of the system -- represent their own interests, rather than attempting to represent their interests for them \cite{Sloan:2020um}. As such, expertise from different domains and fields of life is a practical necessity.

Goal formation in the pre-design phase is contingent on both the context of the larger sociotechnical system for the model and the details of the modeling itself, and the process by which design goals are set out matters because of the power the designers have in shaping real-world outcomes, and their normative consequences. This is not to say that the mechanisms through which these goals are to be accomplished need be determined by the designers before any of the rest of the work of design may begin.  Given the tight bind between the larger goals of the system and the values of the designers, and our observation that the content of the model is important to its normative impact, the reflexive work of goal formalization will often run by necessity concurrently to the process of designing the formal model itself. This will be especially important when it comes to formalizing the goals of the system via optimization functions, null hypotheses for drawing conclusions, etc. 

Perhaps the most critical decision confronting the heterogenous team of modelers, broadly defined, is deciding whether a system should be developed or implemented in the first place. For instance, Malik \cite{Malik:2019ve} argues statistical models be intrinsically marginalizing, but that, ``it doesn't really matter when compared to the choice to use machine learning [in the first place].''  Given that in any situation where a digital system mediates social processes, or is proposed as a mediator, respecting value fidelity requires that the applicability of a technical solution should always be called into question:  there is never an a priori guarantee that an algorithmic system is always going to be the most effective one.

To help modelers decide whether a proposed technical solution is a bad one, there needs to be a process by which a solution is evaluated as suboptimal.  For many computer scientists, there is only one way to answer such questions: try something and compare it to the current baseline. Modelers are thus in something of a double bind: they should not design, build, or implement without considering the consequences of their actions, but it is hard to know the consequences of their actions unless they design, build, and implement.  Selbst et al. \cite{Selbst:2019wf} suggest ``robust conversations between team members'' are central to deciding whether or not to implement a system in the first place. Likewise, Shilton has developed the concept of ``values levers,'' or moments in the design process whereby teams are able to communicate freely about values, ethics and norms \cite{Shilton:2013dd}. In both cases, the implicit point of such conversations is to evaluate a system based on not just what it is intended to do, but also what the system \emph{should} do. Yet there can be no testable prediction for evaluating what a system \emph{should} do, and any designer could try to claim that their system upholds a particular value like fairness by shifting the goalposts to define whatever the system does as fair: a problem made easier because values like fairness are abstract, contested concepts. 

To cope with these challenges, the modeling group should embrace both value fidelity and value contestation as mechanisms through which to reflect on not only what their values are, but how they intend to conceptualize them in the particular domain in which they are building the system. Is the goal of a model equality for the people who use the system? And if so, equality of what, precisely? Scholars have had recent success extrapolating values from existing models, such as Heidari et al.'s work pointing to the philosophical underpinnings of common fairness models in the notion of equality of opportunity \cite{Heidari:2018us}. In turn, some political philosophers have at least endeavoured to produce models explicitly grounded in particular goals grounded in their conceptual versions of fairness as equality of opportunity \cite{Roemer:2012kj}. Though such goals may not initially be conceptualized as a formal model, the point is to have clear goals in the first place, which serve as a groundable hypothesis so that designers can later reflexively test their system for its adherence to the goals set out at the beginning of the project. If there are obvious inconsistencies between normative goals and the means/domain of interest, this is a strong sign the project should not proceed. 
 
In other words, the negative principle of ``know when not to design'' \cite{Baumer:2011uj} is not, as Baumer and Silberman argue, an argument against design and implementation per se: instead, it is a call for technologists to engage in ``critical, reflective dialog about how and why these things are built'' (2274) before they begin the modeling process. Only after testing a hypothesis, however, can the question of whether or not to implement be entirely answered. In the next two sections, we discuss the challenges of actually examining and interrogating such a hypothesis faced during the design of a formal model, and the utility of the reflexive values of accuracy and value legibility in doing so. 

\subsection{Design}

If participants do decide a domain and a system is worth the attempt to model, the next questions must be what and how to model. Recall that formal models are representations of reality, but suffer from a problem of accuracy both of proxies and how proxy data interacts with the mechanisms of the model itself: models often insufficiently proximate to real-world conditions to draw reliable conclusions the best way to situate and operationalize human values in their design.

Value legibility entails articulating and delimiting normative concepts like fairness as discrete mathematical entities. This a necessary first step not only in making that concept formalizable, but also in knowing when not to do so. Modelers can do so through a process of \emph{reflexive discretization}: defining relevant semantics, e.g. defining variables, structures, dynamics, what is and is not observable. Reflexive discretization is always an activity fraught with ethical and moral implications \cite{Barad:2003vh,Bowker:2000vk}.  Such discretization must balance value fidelity, accuracy, and value contestation in translating between observed data, conceptions of that data, particular algorithms, and mathematical objects. Without a high degree of reflexive engagement with all of the above, designers risk inaccuracy or overlooking contested assumptions. For instance, race as an input variable has been an important example of how morally bankrupt concepts underly facially neutral datasets \cite{Benthall:2019dp}.  This legacy makes \emph{how} discretization is done -- with a focus on fidelity, and accuracy -- central for formal modeling.

Attention to value legibility also means modelers must take care as to \emph{what} discretization is performed -- meaning exactly what is being modeled. Without a representation of a particular domain, not only is it harder to ground abstract values (see Section \ref{sec:value_fidelity}), but the use of the model is divorced from how it is used and in what social context (what Selbst et al.\ \cite{Selbst:2019wf} call the ``portability trap'').  Moreover, a model must represent the goals of the system being implemented. Representing the goals of the system in the model is standard operating procedure in computer science and engineering writ large already, but it is worth emphasizing because forming explicit goals are a frequently visited site for contesting values.

Ethical theories, indeed ethical practice, rest on the belief that there is a process or method to analyze a given situation and determine whether it is to be preferred or not.  The information required to do so is what we term the \emph{medium} for values to inform the formal model, and without that information, values cannot be legible in the model. A formal model must have a medium upon which the designers' values operate. Such a medium will depend on the values laid out in the pre-design stage and conceptualized and formed in the design stage.  Values will always operate on some kind of medium.  These challenges parallel what Selbst et. al. \cite{Selbst:2019wf} describe as the ``framing trap,'' or an overly narrow framing of a concept in a formal model. Their proposed solution is to extend the bounding box of a particular discretized concept to be more capacious. In the abstract, this solution to the problem of overly narrow models for complex social phenomena is appealing. It allows for developing a sociotechnical perspective in understanding the social contexts and impacts of machine learning systems.  

However, the application of this move to the process of formal modeling is a prime arena for value contestation, and thus requires reflexive care and an embrace of value legibility and value contestation.  After all, increasing the size and complexity of a model is not necessarily a useful solution to ensuring it captures the particular aspect of sociotechnical complexity of interest. This is an especially acute problem when including incomplete or partial models of complex social processes---such as human social interactions---within one frame of formal abstraction. Human relations are messy, and not easy to model.  Directly modeling the interactions of just two people is an exceedingly difficult challenge.  So increasing the scope of the sociotechnical frame does not necessarily make it easier to model human relations. Rather than focusing on expanding the size of the ``bounding box'' of the model, we advise focusing on value legibility. If any outcome, situation, or process which contravenes the designers' values is not modeled, it has the potential to change the normative valence of the model. This is another area where domain-specific modeling is helpful.  The components in the formal model that make legible the domain can be used as the raw material for modeling particular values.

\subsection{Post-design}

After a model has been developed following the reflexive values described above, modelers should focus on iterative work associated with evaluation, and maintenance, and potential modifications (which may not actually happen after the design of the model).  Of high importance is to test assumptions expressed through the design of the system, including evaluating the accuracy of the model (see Section~\ref{sec:accuracy}):  not only the predictions or conclusions made by the model, but the accuracy of the dynamics assumed or learned via data, and the fidelity of the data itself to the social domain being modeled.

Just as important to these evaluation processes are an ongoing focus on value fidelity and legibility.  Part of the evaluation process is determining the degree to which the proposed system contravenes the designers' values.  User reactions to the system should be of concern, especially if they were not explicitly incorporated into the formal modeling (thus decreasing value legibility).  These user-related activities could include user's attempts to game the system, the kind of user community that the system encourages around it, and figuring out how to meaningfully incorporate user feedback \cite{Sloan:2020um}. 

Assessing these effects is typically performed via a variety of qualitative and quantitative methods: small-scale releases, A/B testing, user interviews, and similar modes of evaluation.  This evaluation process is as susceptible to contravening values as the rest of the design process \cite{Flanagan:2014vd}.  Typical issues include privacy violations, lack of consent to be experimented on, or simply trying out on unsuspecting users an unfinished system that harms them.  Formal modeling of values in this stage of the design process is rare, leading to either poor value legibility or unexamined contested normative assertions. All of these considerations are affected by changing conditions on the ground.  A shifting design context or shifting social domain can affect the normative valence of the system, which means the work of design and the work of evaluation is never done.  Being able to iterate reflexively on a model as it is being developed is key: evaluation methods developed in the context of current VID methods are not necessarily sufficient, especially in the case of algorithmic systems whose second order effects are hard to predict because the systems may be deployed long after modeling is over.

The importance of iteration has long been observed in a variety of kinds of design, but it is typified by the popularity of agile software development, which centers people-focused methodologies via adaptive planning and the participation of concerned stakeholders~\cite{CohenLC04}.  Using agile development over more traditional software development processes has both benefits and drawbacks~\cite{ConboyCWP11,RollandFDS16}, depending on the particular methodology, but the explicit incorporation of values into the design process, exactly because of the contextual nature of values, can make it potentially fertile ground for innovation. Computational modeling more broadly is in some ways still grounded in a waterfall model of software development, moving sequentially from problem space and conceptualization to operationalization, implementation, and iterative testing. This approach contrasts sharply with more free-wheeling methods in agile software developments \cite{Polonetsky:2017vl}. Modeling methods for agile contexts is an ongoing goal of many in the field---and we hope our own agile approach may suggest avenues for future work. 

\section{Discussion} 

Attempting to deal with what Dwork and Mulligan \cite{Dwork:2013ut} call the ``messy reality'' (35) of formal modeling is ``tricky'' indeed \cite{Binns:2018tf} (9). Formal modeling with human values explicitly in mind is challenging, and this analysis represents only a first step of mapping and guiding a rapidly evolving field. The methodology we have presented is largely value-agnostic: our methodological approach to formal modeling can apply to any work that seeks to move from abstract philosophical concepts to concrete technical outputs. As such, it is an approach that can accommodate both cohesive philosophical theories, and particular sets of ad hoc values: moral particularism or philosophical pragmatism are not incompatible with our approach. However, one value of translating a comprehensive political theory is that a theory's many abstract concepts ideally hold internal consistency across a corpus.

We also have no interest in removing either human beings or the wide range of human political theories at hand from the algorithmic fairness loop. We are aware that methods for designing formal models around values have sometimes been critiqued as reifying the idea that formal models are the only language through which algorithmic fairness can be understood. This outcome is certainly not our intention: instead, we hope this methodology is a starting point for computational systems to deploy, for instance, relational notions of equality \cite{Anderson:1999wh}, and shows the benefits of deriving formal models from theories of equality not entirely grounded in material distributions.

A further question is whether mathematical modeling is uniquely suited only to particular ethical frameworks. Consider the case of algorithmic fairness:  despite the large number of definitions of fairness in the machine learning literature, these formalizations codify a relatively narrow range of philosophical definitions around fairness and inequality. There are definitions codifying equality of opportunity as a basis for fairness \cite{Heidari:2018us}; there are also definitions of fairness in this literature formalizing particular conceptions of desert-based ethical frameworks \cite{Dwork:2011vl, JosephKMR16, KusnerLRS17} or of welfare \cite{HeidariFGK18, FishBBFSV19}, but little else.  If you are a strict utilitarian focusing on assigning utilities to actions and then maximizing some function of utility--or believe decisions should be based on individuals' decontextualized `just desserts' and are therefore focused on determining what people deserve based on their feature values--current approaches in machine learning would be uncontroversial \cite{JafariNaimi:2017hg}.

For everybody else, however, it is surely concerning to be wholly reliant on such formal approaches.  For instance, human managers make hiring (and many other) decisions are implicitly using desert-based approaches when they adjudicate between different categories, even without using automated decision systems;  critiques of formalism that find moral fault with this kind of example \cite{Anonymous:2016fr} are thus implicitly addressing not only the use of formal models per se, but the use of those models for adjudicating desert-based ethical theories. The recommendation system for hiring is problematic in this view because it is applying categories to people as a way to decide what they deserve.   

There has been little work seeking to formalize definitions of fairness (let alone definitions of any other value) that draws on the wide array of philosophical explorations of fairness and equality available -- from Kantian or libertarian approaches, to the relational equality of feminist philosophers such as Anderson \cite{Anderson:1999wh}. Is this omission simply because it is easier to express concepts like utility more easily in the language of modeling, or is it because of difficulties particular to these other ethical frameworks?  Variables representing utilities are one thing, but can mathematical structures also appropriately represent and support at least some of the intricacies of social interactions that underpin values such as justice or human flourishing?  Understanding how diverse normative frameworks might be expressed through mathematical modeling is an under-explored area of algorithmic fairness, one in which reflexive design practices described above are vital: without reflexivity, it is a challenge both to answer the question of whether formal modeling is entirely unsuited to some definitions of values, and if not, how formal models representing such values should be designed.

Finally, given that reflexive iteration is an important value for our method, the method itself will also benefit from refinement and application. This method springs from our own efforts at translating values into formal models, and is intended to guide modelers in practice -- it would benefit from further application in pilot projects and further studies. More broadly, future research should explore the degree to which ethical frameworks and other value systems can be explored appropriately in modeling, outside of the relatively limited approaches used in the computer science literature to date, and to the degree to which formal approaches can be used to model the socio-technical systems themselves. This research is beneficial not just for applied approaches, but to the theoretical community interested in modeling algorithmic fairness -- these approaches have the potential to suggest new formal problems, as well as challenge current applications in the field. 

We are mindful of Deleuze's prescient observation that digital technologies merely ``express the social forms capable of producing them and making use of them'' \cite{Deleuze:1990uk}. In other words, for all the ways in which digital systems embody values, the effects of these systems on people stem from the material conditions, and social relations in the world at large.  Modeling that takes these concerns into account reflexively in the design process will, we believe, serve to support broader efforts to understand and ameliorate the ways in which digital systems contribute to shaping private and public life.

\bibliographystyle{ACM-Reference-Format}
\balance
\bibliography{refs}


\begin{thebibliography}{00}


\ifx \showCODEN    \undefined \def \showCODEN     #1{\unskip}     \fi
\ifx \showDOI      \undefined \def \showDOI       #1{#1}\fi
\ifx \showISBNx    \undefined \def \showISBNx     #1{\unskip}     \fi
\ifx \showISBNxiii \undefined \def \showISBNxiii  #1{\unskip}     \fi
\ifx \showISSN     \undefined \def \showISSN      #1{\unskip}     \fi
\ifx \showLCCN     \undefined \def \showLCCN      #1{\unskip}     \fi
\ifx \shownote     \undefined \def \shownote      #1{#1}          \fi
\ifx \showarticletitle \undefined \def \showarticletitle #1{#1}   \fi
\ifx \showURL      \undefined \def \showURL       {\relax}        \fi
\providecommand\bibfield[2]{#2}
\providecommand\bibinfo[2]{#2}
\providecommand\natexlab[1]{#1}
\providecommand\showeprint[2][]{arXiv:#2}

\bibitem[\protect\citeauthoryear{Abebe, Barocas, Kleinberg, Karen, Raghavan,
  and Robinson}{Abebe et~al\mbox{.}}{2020}]%
        {Abebe:2020kr}
\bibfield{author}{\bibinfo{person}{Rediet Abebe}, \bibinfo{person}{Solon
  Barocas}, \bibinfo{person}{Jon Kleinberg}, \bibinfo{person}{Levy Karen},
  \bibinfo{person}{Manish Raghavan}, {and} \bibinfo{person}{David~G Robinson}.}
  \bibinfo{year}{2020}\natexlab{}.
\newblock \showarticletitle{{Roles for Computing in Social Change}}. In
  \bibinfo{booktitle}{{\em FAccT '20}}. \bibinfo{address}{Barcelona},
  \bibinfo{pages}{1--9}.
\newblock


\bibitem[\protect\citeauthoryear{Abebe and Goldner}{Abebe and Goldner}{2018}]%
        {Abebe:2018jc}
\bibfield{author}{\bibinfo{person}{Rediet Abebe} {and} \bibinfo{person}{Kira
  Goldner}.} \bibinfo{year}{2018}\natexlab{}.
\newblock \showarticletitle{{Mechanism design for social good}}.
\newblock \bibinfo{journal}{{\em AI Matters\/}} \bibinfo{volume}{4},
  \bibinfo{number}{3} (\bibinfo{date}{Oct.} \bibinfo{year}{2018}),
  \bibinfo{pages}{27--34}.
\newblock


\bibitem[\protect\citeauthoryear{Agre}{Agre}{1997a}]%
        {Agre:1997tl}
\bibfield{author}{\bibinfo{person}{Philip~E. Agre}.}
  \bibinfo{year}{1997}\natexlab{a}.
\newblock \bibinfo{booktitle}{{\em {Computation and human experience}}}.
\newblock \bibinfo{publisher}{Cambridge University Press},
  \bibinfo{address}{Cambridge, UK}.
\newblock


\bibitem[\protect\citeauthoryear{Agre}{Agre}{1997b}]%
        {Agre:1997ts}
\bibfield{author}{\bibinfo{person}{Philip~E. Agre}.}
  \bibinfo{year}{1997}\natexlab{b}.
\newblock \showarticletitle{{Toward a Critical Technical Practice: Lessons
  Learned in Trying to Reform AI}}.
\newblock In \bibinfo{booktitle}{{\em Social Science, Technical Systems and
  Cooperative Work: The Great Divide}},
  \bibfield{editor}{\bibinfo{person}{Geoffrey~C. Bowker}, \bibinfo{person}{Les
  Gasser}, \bibinfo{person}{Susan~Leigh Star}, {and} \bibinfo{person}{Bill
  Turner}} (Eds.). \bibinfo{address}{Hillsdale, NJ}, \bibinfo{pages}{131--158}.
\newblock


\bibitem[\protect\citeauthoryear{Ananny}{Ananny}{2015}]%
        {Ananny:2015fn}
\bibfield{author}{\bibinfo{person}{Mike Ananny}.}
  \bibinfo{year}{2015}\natexlab{}.
\newblock \showarticletitle{{Toward an Ethics of Algorithms}}.
\newblock \bibinfo{journal}{{\em Science, Technology, {\&} Human Values\/}}
  \bibinfo{volume}{41}, \bibinfo{number}{1} (\bibinfo{year}{2015}),
  \bibinfo{pages}{93--117}.
\newblock


\bibitem[\protect\citeauthoryear{Anderson}{Anderson}{1999}]%
        {Anderson:1999wh}
\bibfield{author}{\bibinfo{person}{Elizabeth~S Anderson}.}
  \bibinfo{year}{1999}\natexlab{}.
\newblock \showarticletitle{{What Is the Point of Equality?}}
\newblock \bibinfo{journal}{{\em Ethics\/}} \bibinfo{volume}{109},
  \bibinfo{number}{2} (\bibinfo{date}{Jan.} \bibinfo{year}{1999}),
  \bibinfo{pages}{287--337}.
\newblock


\bibitem[\protect\citeauthoryear{Barad}{Barad}{2003}]%
        {Barad:2003vh}
\bibfield{author}{\bibinfo{person}{Karen Barad}.}
  \bibinfo{year}{2003}\natexlab{}.
\newblock \showarticletitle{{Posthumanist Performativity: Toward an
  Understanding of How Matter Comes to Matter}}.
\newblock \bibinfo{journal}{{\em Signs\/}} \bibinfo{volume}{28},
  \bibinfo{number}{3} (\bibinfo{date}{April} \bibinfo{year}{2003}),
  \bibinfo{pages}{801--831}.
\newblock


\bibitem[\protect\citeauthoryear{Barocas and Selbst}{Barocas and
  Selbst}{2016}]%
        {Anonymous:2016fr}
\bibfield{author}{\bibinfo{person}{Solon Barocas} {and}
  \bibinfo{person}{Andrew~D. Selbst}.} \bibinfo{year}{2016}\natexlab{}.
\newblock \showarticletitle{{Big Data{\textquoteright}s Disparate Impact}}.
\newblock \bibinfo{journal}{{\em California Law Review\/}}
  \bibinfo{volume}{104} (\bibinfo{year}{2016}), \bibinfo{pages}{671--732}.
\newblock


\bibitem[\protect\citeauthoryear{Baumer and Silberman}{Baumer and
  Silberman}{2011}]%
        {Baumer:2011uj}
\bibfield{author}{\bibinfo{person}{Eric P.~S. Baumer} {and}
  \bibinfo{person}{M~Six Silberman}.} \bibinfo{year}{2011}\natexlab{}.
\newblock \showarticletitle{{When the Implication Is Not to Design
  (Technology)}}. In \bibinfo{booktitle}{{\em CHI 2011}}.
  \bibinfo{address}{Vancouver, Canada}, \bibinfo{pages}{2271--2274}.
\newblock


\bibitem[\protect\citeauthoryear{Benthall and Haynes}{Benthall and
  Haynes}{2019}]%
        {Benthall:2019dp}
\bibfield{author}{\bibinfo{person}{Sebastian Benthall} {and}
  \bibinfo{person}{Bruce~D Haynes}.} \bibinfo{year}{2019}\natexlab{}.
\newblock \showarticletitle{{Racial categories in machine learning}}. In
  \bibinfo{booktitle}{{\em FAT 2019}}. \bibinfo{address}{New York, New York,
  USA}, \bibinfo{pages}{289--298}.
\newblock


\bibitem[\protect\citeauthoryear{Binns}{Binns}{2018}]%
        {Binns:2018tf}
\bibfield{author}{\bibinfo{person}{Rueben Binns}.}
  \bibinfo{year}{2018}\natexlab{}.
\newblock \showarticletitle{{Fairness in Machine Learning: Lessons from
  Political Philosophy}}.
\newblock \bibinfo{journal}{{\em Proceedings of Machine Learning Research\/}}
  \bibinfo{volume}{81} (\bibinfo{year}{2018}), \bibinfo{pages}{1--11}.
\newblock


\bibitem[\protect\citeauthoryear{Bowker and Star}{Bowker and Star}{2000}]%
        {Bowker:2000vk}
\bibfield{author}{\bibinfo{person}{Geoffrey~C. Bowker} {and}
  \bibinfo{person}{Susan~Leigh Star}.} \bibinfo{year}{2000}\natexlab{}.
\newblock \bibinfo{booktitle}{{\em {Sorting Things Out: Classification and Its
  Consequences (Inside Technology)}}}.
\newblock \bibinfo{publisher}{The MIT Press}, \bibinfo{address}{Cambridge, MA}.
\newblock


\bibitem[\protect\citeauthoryear{Cohen, Lindvall, and Costa}{Cohen
  et~al\mbox{.}}{2004}]%
        {CohenLC04}
\bibfield{author}{\bibinfo{person}{David Cohen}, \bibinfo{person}{Mikael
  Lindvall}, {and} \bibinfo{person}{Patricia Costa}.}
  \bibinfo{year}{2004}\natexlab{}.
\newblock \showarticletitle{An introduction to agile methods}.
\newblock \bibinfo{journal}{{\em Adv. Comput.\/}}  \bibinfo{volume}{62}
  (\bibinfo{year}{2004}), \bibinfo{pages}{1--66}.
\newblock
\showDOI{%
\url{https://doi.org/10.1016/S0065-2458(03)62001-2}}


\bibitem[\protect\citeauthoryear{Conboy, Coyle, Wang, and Pikkarainen}{Conboy
  et~al\mbox{.}}{2011}]%
        {ConboyCWP11}
\bibfield{author}{\bibinfo{person}{Kieran Conboy}, \bibinfo{person}{Sharon
  Coyle}, \bibinfo{person}{Xiaofeng Wang}, {and} \bibinfo{person}{Minna
  Pikkarainen}.} \bibinfo{year}{2011}\natexlab{}.
\newblock \showarticletitle{People over Process: Key Challenges in Agile
  Development}.
\newblock \bibinfo{journal}{{\em {IEEE} Softw.\/}} \bibinfo{volume}{28},
  \bibinfo{number}{4} (\bibinfo{year}{2011}), \bibinfo{pages}{48--57}.
\newblock
\showDOI{%
\url{https://doi.org/10.1109/MS.2010.132}}


\bibitem[\protect\citeauthoryear{Corbett-Davies and Goel}{Corbett-Davies and
  Goel}{2018}]%
        {CorbettDavies:2018ud}
\bibfield{author}{\bibinfo{person}{Sam Corbett-Davies} {and}
  \bibinfo{person}{Sharad Goel}.} \bibinfo{year}{2018}\natexlab{}.
\newblock \showarticletitle{{The Measure and Mismeasure of Fairness: A Critical
  Review of Fair Machine Learning}}.
\newblock  (\bibinfo{date}{Sept.} \bibinfo{year}{2018}),
  \bibinfo{pages}{1--25}.
\newblock


\bibitem[\protect\citeauthoryear{Deleuze}{Deleuze}{1990}]%
        {Deleuze:1990uk}
\bibfield{author}{\bibinfo{person}{Gilles Deleuze}.}
  \bibinfo{year}{1990}\natexlab{}.
\newblock \showarticletitle{{Postscript on Control Societies}}.
\newblock In \bibinfo{booktitle}{{\em Negotiations, 1972-1990}}.
  \bibinfo{publisher}{Columbia University Press}, \bibinfo{address}{New York},
  \bibinfo{pages}{177--182}.
\newblock


\bibitem[\protect\citeauthoryear{Dwork, Hardt, Pitassi, Reingold, and
  Zemel}{Dwork et~al\mbox{.}}{2011}]%
        {Dwork:2011vl}
\bibfield{author}{\bibinfo{person}{Cynthia Dwork}, \bibinfo{person}{Moritz
  Hardt}, \bibinfo{person}{Toniann Pitassi}, \bibinfo{person}{Omer Reingold},
  {and} \bibinfo{person}{Richard Zemel}.} \bibinfo{year}{2011}\natexlab{}.
\newblock \showarticletitle{{Fairness Through Awareness}}.
\newblock  (\bibinfo{date}{Nov.} \bibinfo{year}{2011}), \bibinfo{pages}{1--24}.
\newblock


\bibitem[\protect\citeauthoryear{Dwork and Mulligan}{Dwork and
  Mulligan}{2013}]%
        {Dwork:2013ut}
\bibfield{author}{\bibinfo{person}{Cynthia Dwork} {and}
  \bibinfo{person}{Deirdre~K. Mulligan}.} \bibinfo{year}{2013}\natexlab{}.
\newblock \showarticletitle{{It's Not Privacy, and It's Not Fair}}.
\newblock \bibinfo{journal}{{\em Stanford Law Review Online\/}}
  \bibinfo{volume}{66} (\bibinfo{date}{Sept.} \bibinfo{year}{2013}),
  \bibinfo{pages}{35--40}.
\newblock


\bibitem[\protect\citeauthoryear{Elliott and McKaughan}{Elliott and
  McKaughan}{2014}]%
        {Elliott:2014uq}
\bibfield{author}{\bibinfo{person}{Kevin~C. Elliott} {and}
  \bibinfo{person}{Daniel~J. McKaughan}.} \bibinfo{year}{2014}\natexlab{}.
\newblock \showarticletitle{{Nonepistemic Values and the Multiple Goals of
  Science}}.
\newblock \bibinfo{journal}{{\em Philosophy of Science\/}}
  \bibinfo{volume}{81} (\bibinfo{date}{Jan.} \bibinfo{year}{2014}),
  \bibinfo{pages}{1--21}.
\newblock


\bibitem[\protect\citeauthoryear{Eubanks}{Eubanks}{2018a}]%
        {Eubanks:2018wv}
\bibfield{author}{\bibinfo{person}{Virgina Eubanks}.}
  \bibinfo{year}{2018}\natexlab{a}.
\newblock \bibinfo{booktitle}{{\em {Automating Inequality: How High-Tech Tools
  Profile, Police, and Punish the Poor}}}.
\newblock \bibinfo{publisher}{St. Martin's Press}, \bibinfo{address}{New York}.
\newblock


\bibitem[\protect\citeauthoryear{Eubanks}{Eubanks}{2018b}]%
        {Eubanks:2018uy}
\bibfield{author}{\bibinfo{person}{Virgina Eubanks}.}
  \bibinfo{year}{2018}\natexlab{b}.
\newblock \bibinfo{title}{{The Digital Poorhouse}}.
\newblock   (\bibinfo{date}{Jan.} \bibinfo{year}{2018}).
\newblock
\showURL{%
\url{https://harpers.org/archive/2018/01/the-digital-poorhouse/}}


\bibitem[\protect\citeauthoryear{Fish, Bashardoust, danah boyd, Friedler,
  Scheidegger, and Venkatasubramanian}{Fish et~al\mbox{.}}{2019}]%
        {FishBBFSV19}
\bibfield{author}{\bibinfo{person}{Benjamin Fish}, \bibinfo{person}{Ashkan
  Bashardoust}, \bibinfo{person}{danah boyd}, \bibinfo{person}{Sorelle~A.
  Friedler}, \bibinfo{person}{Carlos Scheidegger}, {and}
  \bibinfo{person}{Suresh Venkatasubramanian}.}
  \bibinfo{year}{2019}\natexlab{}.
\newblock \showarticletitle{Gaps in Information Access in Social Networks}. In
  \bibinfo{booktitle}{{\em The World Wide Web Conference, {WWW} 2019, San
  Francisco, CA, USA, May 13-17, 2019}}. \bibinfo{pages}{480--490}.
\newblock


\bibitem[\protect\citeauthoryear{Flanagan, Howe, and Nissenbaum}{Flanagan
  et~al\mbox{.}}{2008}]%
        {Flanagan:2008wx}
\bibfield{author}{\bibinfo{person}{Mary Flanagan}, \bibinfo{person}{Daniel~C
  Howe}, {and} \bibinfo{person}{Helen Nissenbaum}.}
  \bibinfo{year}{2008}\natexlab{}.
\newblock \showarticletitle{{Embodying Values in Technology: Theory and
  Practice}}.
\newblock In \bibinfo{booktitle}{{\em Information Technology and Moral
  Philosophy}}, \bibfield{editor}{\bibinfo{person}{Jeroen van~den Hoeven} {and}
  \bibinfo{person}{John Weckert}} (Eds.). \bibinfo{publisher}{Cambridge
  University Press}, \bibinfo{address}{Cambridge, UK},
  \bibinfo{pages}{322--353}.
\newblock


\bibitem[\protect\citeauthoryear{Flanagan and Nissenbaum}{Flanagan and
  Nissenbaum}{2014}]%
        {Flanagan:2014vd}
\bibfield{author}{\bibinfo{person}{Mary Flanagan} {and} \bibinfo{person}{Helen
  Nissenbaum}.} \bibinfo{year}{2014}\natexlab{}.
\newblock \bibinfo{booktitle}{{\em {Values at Play in Digital Games}}}.
\newblock \bibinfo{publisher}{The MIT Press}, \bibinfo{address}{Cambridge, MA}.
\newblock


\bibitem[\protect\citeauthoryear{Fleischmann, Hui, and Wallace}{Fleischmann
  et~al\mbox{.}}{2016}]%
        {Fleischmann:2016ga}
\bibfield{author}{\bibinfo{person}{Kenneth~R Fleischmann},
  \bibinfo{person}{Cindy Hui}, {and} \bibinfo{person}{William~A. Wallace}.}
  \bibinfo{year}{2016}\natexlab{}.
\newblock \showarticletitle{{The Societal Responsibilities of Computational
  Modelers: Human Values and Professional Codes of Ethics}}.
\newblock \bibinfo{journal}{{\em Journal of the Association for Information
  Science and Technology\/}} \bibinfo{volume}{68}, \bibinfo{number}{3}
  (\bibinfo{date}{June} \bibinfo{year}{2016}), \bibinfo{pages}{543--552}.
\newblock


\bibitem[\protect\citeauthoryear{Fleischmann and Wallace}{Fleischmann and
  Wallace}{2010}]%
        {Fleischmann:2010tj}
\bibfield{author}{\bibinfo{person}{Kenneth~R. Fleischmann} {and}
  \bibinfo{person}{William~A. Wallace}.} \bibinfo{year}{2010}\natexlab{}.
\newblock \showarticletitle{{Value Conflicts in Computational Modeling}}.
\newblock \bibinfo{journal}{{\em Computer\/}}  \bibinfo{volume}{43}
  (\bibinfo{date}{July} \bibinfo{year}{2010}), \bibinfo{pages}{57--63}.
\newblock


\bibitem[\protect\citeauthoryear{Fleischmann, Wallace, and Grimes}{Fleischmann
  et~al\mbox{.}}{2011}]%
        {Fleischmann:2011ws}
\bibfield{author}{\bibinfo{person}{Kenneth~R. Fleischmann},
  \bibinfo{person}{William~A. Wallace}, {and} \bibinfo{person}{Justin~M.
  Grimes}.} \bibinfo{year}{2011}\natexlab{}.
\newblock \showarticletitle{{How Values Can Reduce Conflicts in the Design
  Process: Results from a Multi-Site Mixed-Method Field Study}}. In
  \bibinfo{booktitle}{{\em ASIST 2011}}. \bibinfo{pages}{1--10}.
\newblock


\bibitem[\protect\citeauthoryear{Fortnow and Homer}{Fortnow and Homer}{2003}]%
        {FortnowH03}
\bibfield{author}{\bibinfo{person}{Lance Fortnow} {and} \bibinfo{person}{Steven
  Homer}.} \bibinfo{year}{2003}\natexlab{}.
\newblock \showarticletitle{A Short History of Computational Complexity}.
\newblock \bibinfo{journal}{{\em Bull. {EATCS}\/}}  \bibinfo{volume}{80}
  (\bibinfo{year}{2003}), \bibinfo{pages}{95--133}.
\newblock


\bibitem[\protect\citeauthoryear{Fourcade}{Fourcade}{2016}]%
        {Fourcade:2016ge}
\bibfield{author}{\bibinfo{person}{Marion Fourcade}.}
  \bibinfo{year}{2016}\natexlab{}.
\newblock \showarticletitle{{Ordinalization}}.
\newblock \bibinfo{journal}{{\em Sociological Theory\/}} \bibinfo{volume}{34},
  \bibinfo{number}{3} (\bibinfo{date}{Sept.} \bibinfo{year}{2016}),
  \bibinfo{pages}{175--195}.
\newblock


\bibitem[\protect\citeauthoryear{Friedland and Yamauchi}{Friedland and
  Yamauchi}{2011}]%
        {Friedland:2011bm}
\bibfield{author}{\bibinfo{person}{Barton Friedland} {and}
  \bibinfo{person}{Yutaka Yamauchi}.} \bibinfo{year}{2011}\natexlab{}.
\newblock \showarticletitle{{Reflexive Design Thinking: Putting More Human in
  Human-Centered Practices}}.
\newblock \bibinfo{journal}{{\em Interactions\/}} \bibinfo{volume}{18},
  \bibinfo{number}{2} (\bibinfo{year}{2011}), \bibinfo{pages}{66--71}.
\newblock


\bibitem[\protect\citeauthoryear{Friedman and Hendry}{Friedman and
  Hendry}{2019}]%
        {Friedman:2019vu}
\bibfield{author}{\bibinfo{person}{Batya Friedman} {and}
  \bibinfo{person}{David~G. Hendry}.} \bibinfo{year}{2019}\natexlab{}.
\newblock \bibinfo{booktitle}{{\em {Value Sensitive Design}}}.
\newblock \bibinfo{publisher}{The MIT Press}, \bibinfo{address}{Cambridge, MA}.
\newblock


\bibitem[\protect\citeauthoryear{Friedman, Kahn, and Borning}{Friedman
  et~al\mbox{.}}{2006}]%
        {Friedman:2006vb}
\bibfield{author}{\bibinfo{person}{Batya Friedman}, \bibinfo{person}{Peter~H
  Kahn}, {and} \bibinfo{person}{Alan Borning}.}
  \bibinfo{year}{2006}\natexlab{}.
\newblock \showarticletitle{{Value Sensitive Design and Information Systems}}.
\newblock In \bibinfo{booktitle}{{\em Human-Computer Interaction in Management
  Information Systems: Foundations}},
  \bibfield{editor}{\bibinfo{person}{B~Schneiderman}, \bibinfo{person}{Ping
  Zhang}, {and} \bibinfo{person}{D~Galletta}} (Eds.). \bibinfo{publisher}{M.E.
  Sharpe, Inc.}, \bibinfo{address}{New York}, \bibinfo{pages}{348--372}.
\newblock


\bibitem[\protect\citeauthoryear{Friedman and Nissenbaum}{Friedman and
  Nissenbaum}{1996}]%
        {Friedman:1996tm}
\bibfield{author}{\bibinfo{person}{Batya Friedman} {and} \bibinfo{person}{Helen
  Nissenbaum}.} \bibinfo{year}{1996}\natexlab{}.
\newblock \showarticletitle{{Bias in Computer Systems}}.
\newblock \bibinfo{journal}{{\em ACM Transactions on Information Systems\/}}
  \bibinfo{volume}{14}, \bibinfo{number}{3} (\bibinfo{date}{Sept.}
  \bibinfo{year}{1996}), \bibinfo{pages}{330--347}.
\newblock


\bibitem[\protect\citeauthoryear{Green and Viljoen}{Green and Viljoen}{2020}]%
        {Green:2020boa}
\bibfield{author}{\bibinfo{person}{Ben Green} {and} \bibinfo{person}{Salom{\'e}
  Viljoen}.} \bibinfo{year}{2020}\natexlab{}.
\newblock \showarticletitle{{Algorithmic Realism}}. In \bibinfo{booktitle}{{\em
  FAT* '20: Conference on Fairness, Accountability, and Transparency}}.
  \bibinfo{publisher}{ACM}, \bibinfo{address}{New York, NY, USA},
  \bibinfo{pages}{19--31}.
\newblock


\bibitem[\protect\citeauthoryear{G{\"u}rses and Hoboken}{G{\"u}rses and
  Hoboken}{2017}]%
        {Polonetsky:2017vl}
\bibfield{author}{\bibinfo{person}{Seda G{\"u}rses} {and}
  \bibinfo{person}{Joris~van Hoboken}.} \bibinfo{year}{2017}\natexlab{}.
\newblock \showarticletitle{{Privacy After the Agile Turn}}.
\newblock In \bibinfo{booktitle}{{\em Cambridge Handbook of Consumer Privacy}},
  \bibfield{editor}{\bibinfo{person}{Jules Polonetsky}, \bibinfo{person}{Omer
  Tene}, {and} \bibinfo{person}{Evan Selinger}} (Eds.).
  \bibinfo{address}{CambrIdge, UK}.
\newblock


\bibitem[\protect\citeauthoryear{Haraway}{Haraway}{1988}]%
        {Haraway:1988us}
\bibfield{author}{\bibinfo{person}{Donna~J Haraway}.}
  \bibinfo{year}{1988}\natexlab{}.
\newblock \showarticletitle{{Situated Knowledges}}.
\newblock \bibinfo{journal}{{\em Feminist Studies\/}} \bibinfo{volume}{14},
  \bibinfo{number}{3} (\bibinfo{date}{Oct.} \bibinfo{year}{1988}),
  \bibinfo{pages}{575--599}.
\newblock


\bibitem[\protect\citeauthoryear{Heidari, Ferrari, Gummadi, and Krause}{Heidari
  et~al\mbox{.}}{2018a}]%
        {HeidariFGK18}
\bibfield{author}{\bibinfo{person}{Hoda Heidari}, \bibinfo{person}{Claudio
  Ferrari}, \bibinfo{person}{Krishna~P. Gummadi}, {and}
  \bibinfo{person}{Andreas Krause}.} \bibinfo{year}{2018}\natexlab{a}.
\newblock \showarticletitle{Fairness Behind a Veil of Ignorance: {A} Welfare
  Analysis for Automated Decision Making}. In \bibinfo{booktitle}{{\em Advances
  in Neural Information Processing Systems 31: Annual Conference on Neural
  Information Processing Systems 2018, NeurIPS 2018, 3-8 December 2018,
  Montr{\'{e}}al, Canada}}. \bibinfo{pages}{1273--1283}.
\newblock


\bibitem[\protect\citeauthoryear{Heidari, Loi, Gummadi, and Krause}{Heidari
  et~al\mbox{.}}{2018b}]%
        {Heidari:2018us}
\bibfield{author}{\bibinfo{person}{Hoda Heidari}, \bibinfo{person}{Michele
  Loi}, \bibinfo{person}{Krishna~P. Gummadi}, {and} \bibinfo{person}{Andreas
  Krause}.} \bibinfo{year}{2018}\natexlab{b}.
\newblock \showarticletitle{{A Moral Framework for Understanding of Fair ML
  through Economic Models of Equality of Opportunity}}.
\newblock  (\bibinfo{date}{Sept.} \bibinfo{year}{2018}),
  \bibinfo{pages}{1--13}.
\newblock
\showeprint{1809.03400}


\bibitem[\protect\citeauthoryear{Hoffmann}{Hoffmann}{2019}]%
        {Hoffmann:2019hr}
\bibfield{author}{\bibinfo{person}{Anna~Lauren Hoffmann}.}
  \bibinfo{year}{2019}\natexlab{}.
\newblock \showarticletitle{{Where fairness fails: data, algorithms, and the
  limits of antidiscrimination discourse}}.
\newblock \bibinfo{journal}{{\em Information, Communication {\&} Society\/}}
  \bibinfo{volume}{22}, \bibinfo{number}{7} (\bibinfo{date}{May}
  \bibinfo{year}{2019}), \bibinfo{pages}{900--915}.
\newblock


\bibitem[\protect\citeauthoryear{Howe and Nissenbaum}{Howe and
  Nissenbaum}{2009}]%
        {Howe:2009uj}
\bibfield{author}{\bibinfo{person}{Daniel~C Howe} {and} \bibinfo{person}{Helen
  Nissenbaum}.} \bibinfo{year}{2009}\natexlab{}.
\newblock \showarticletitle{{TrackMeNot: Resisting Surveillance in Web
  Search}}.
\newblock In \bibinfo{booktitle}{{\em Lessons from the Identity Trail
  Anonymity, Privacy and Identity in a Networked Society}},
  \bibfield{editor}{\bibinfo{person}{Ian Kerr}, \bibinfo{person}{Valerie
  Steeves}, {and} \bibinfo{person}{Carole Lucock}} (Eds.).
  \bibinfo{publisher}{Oxford University Press}, \bibinfo{address}{Oxford and
  New York}, \bibinfo{pages}{417--436}.
\newblock


\bibitem[\protect\citeauthoryear{Hughes}{Hughes}{1987}]%
        {Hughes:1987tv}
\bibfield{author}{\bibinfo{person}{Thomas~P Hughes}.}
  \bibinfo{year}{1987}\natexlab{}.
\newblock \showarticletitle{{The Evolution of Large Technological Systems}}.
\newblock In \bibinfo{booktitle}{{\em The Social Construction of Technological
  Systems: New Directions in the Sociology and History of Technology}},
  \bibfield{editor}{\bibinfo{person}{Wiebe~E. Bijker},
  \bibinfo{person}{Thomas~P Hughes}, {and} \bibinfo{person}{Trevor Pinch}}
  (Eds.). \bibinfo{address}{Cambridge, MA}, \bibinfo{pages}{51--82}.
\newblock


\bibitem[\protect\citeauthoryear{Hutchinson and Mitchell}{Hutchinson and
  Mitchell}{2019}]%
        {Hutchinson:2019ce}
\bibfield{author}{\bibinfo{person}{Ben Hutchinson} {and}
  \bibinfo{person}{Margaret Mitchell}.} \bibinfo{year}{2019}\natexlab{}.
\newblock \showarticletitle{{50 Years of Test (Un)fairness}}. In
  \bibinfo{booktitle}{{\em Conference on Fairness, Accountability, and
  Transparency 2019}}. \bibinfo{publisher}{ACM Press}, \bibinfo{address}{New
  York, New York, USA}, \bibinfo{pages}{49--58}.
\newblock


\bibitem[\protect\citeauthoryear{Jacobs and Wallach}{Jacobs and
  Wallach}{2021}]%
        {Jacobs:2019vx}
\bibfield{author}{\bibinfo{person}{Abigail~Z. Jacobs} {and}
  \bibinfo{person}{Hanna Wallach}.} \bibinfo{year}{2021}\natexlab{}.
\newblock \showarticletitle{Measurement and Fairness}. In
  \bibinfo{booktitle}{{\em FAccT '21: 2021 {ACM} Conference on Fairness,
  Accountability, and Transparency, Virtual Event / Toronto, Canada, March
  3-10, 2021}}. \bibinfo{pages}{375--385}.
\newblock


\bibitem[\protect\citeauthoryear{JafariNaimi}{JafariNaimi}{2017}]%
        {JafariNaimi:2017hg}
\bibfield{author}{\bibinfo{person}{Nassim JafariNaimi}.}
  \bibinfo{year}{2017}\natexlab{}.
\newblock \showarticletitle{{Our Bodies in the Trolley{\textquoteright}s Path,
  or Why Self-driving Cars Must *Not* Be Programmed to Kill}}.
\newblock \bibinfo{journal}{{\em Science, Technology, {\&} Human Values\/}}
  \bibinfo{volume}{5}, \bibinfo{number}{2} (\bibinfo{date}{Nov.}
  \bibinfo{year}{2017}), \bibinfo{pages}{016224391771894}.
\newblock


\bibitem[\protect\citeauthoryear{JafariNaimi, Nathan, and
  Hargraves}{JafariNaimi et~al\mbox{.}}{2015}]%
        {JafariNaimi:2015en}
\bibfield{author}{\bibinfo{person}{Nassim JafariNaimi}, \bibinfo{person}{Lisa
  Nathan}, {and} \bibinfo{person}{Ian Hargraves}.}
  \bibinfo{year}{2015}\natexlab{}.
\newblock \showarticletitle{{Values as Hypotheses: Design, Inquiry, and the
  Service of Values}}.
\newblock \bibinfo{journal}{{\em Design Issues\/}} \bibinfo{volume}{31},
  \bibinfo{number}{4} (\bibinfo{date}{Oct.} \bibinfo{year}{2015}),
  \bibinfo{pages}{91--104}.
\newblock


\bibitem[\protect\citeauthoryear{Joseph, Kearns, Morgenstern, and Roth}{Joseph
  et~al\mbox{.}}{2016}]%
        {JosephKMR16}
\bibfield{author}{\bibinfo{person}{Matthew Joseph}, \bibinfo{person}{Michael~J.
  Kearns}, \bibinfo{person}{Jamie~H. Morgenstern}, {and} \bibinfo{person}{Aaron
  Roth}.} \bibinfo{year}{2016}\natexlab{}.
\newblock \showarticletitle{Fairness in Learning: Classic and Contextual
  Bandits}. In \bibinfo{booktitle}{{\em Advances in Neural Information
  Processing Systems 29: Annual Conference on Neural Information Processing
  Systems 2016, December 5-10, 2016, Barcelona, Spain}}.
  \bibinfo{pages}{325--333}.
\newblock


\bibitem[\protect\citeauthoryear{Kearns, Neel, Roth, and Wu}{Kearns
  et~al\mbox{.}}{2018}]%
        {Kearns:2018wp}
\bibfield{author}{\bibinfo{person}{Michael Kearns}, \bibinfo{person}{Seth
  Neel}, \bibinfo{person}{Aaron Roth}, {and} \bibinfo{person}{Zhiwei~Steven
  Wu}.} \bibinfo{year}{2018}\natexlab{}.
\newblock \showarticletitle{{Preventing Fairness Gerrymandering: Auditing and
  Learning for Subgroup Fairness}}.
\newblock \bibinfo{journal}{{\em arXiv\/}} (\bibinfo{date}{Dec.}
  \bibinfo{year}{2018}), \bibinfo{pages}{1--34}.
\newblock


\bibitem[\protect\citeauthoryear{Keyes}{Keyes}{2019}]%
        {Keyes:2019vi}
\bibfield{author}{\bibinfo{person}{Os Keyes}.} \bibinfo{year}{2019}\natexlab{}.
\newblock \bibinfo{title}{{Counting the Countless}}.
\newblock   (\bibinfo{date}{April} \bibinfo{year}{2019}).
\newblock
\showURL{%
\url{https://reallifemag.com/counting-the-countless/}}


\bibitem[\protect\citeauthoryear{Kleinberg, Ludwig, and Mullainathan}{Kleinberg
  et~al\mbox{.}}{2016}]%
        {kleinberg2016guide}
\bibfield{author}{\bibinfo{person}{Jon Kleinberg}, \bibinfo{person}{Jens
  Ludwig}, {and} \bibinfo{person}{Sendhil Mullainathan}.}
  \bibinfo{year}{2016}\natexlab{}.
\newblock \showarticletitle{A Guide to Solving Social Problems with Machine
  Learning}.
\newblock \bibinfo{journal}{{\em Harvard Business Review\/}}
  (\bibinfo{year}{2016}).
\newblock


\bibitem[\protect\citeauthoryear{Knobel and Bowker}{Knobel and Bowker}{2011}]%
        {Knobel:2011hza}
\bibfield{author}{\bibinfo{person}{Cory Knobel} {and}
  \bibinfo{person}{Geoffrey~C. Bowker}.} \bibinfo{year}{2011}\natexlab{}.
\newblock \showarticletitle{{Values in design}}.
\newblock \bibinfo{journal}{{\it Commun. ACM}} \bibinfo{volume}{54},
  \bibinfo{number}{7} (\bibinfo{date}{July} \bibinfo{year}{2011}),
  \bibinfo{pages}{26--4}.
\newblock


\bibitem[\protect\citeauthoryear{Kusner, Loftus, Russell, and Silva}{Kusner
  et~al\mbox{.}}{2017}]%
        {KusnerLRS17}
\bibfield{author}{\bibinfo{person}{Matt~J. Kusner}, \bibinfo{person}{Joshua~R.
  Loftus}, \bibinfo{person}{Chris Russell}, {and} \bibinfo{person}{Ricardo
  Silva}.} \bibinfo{year}{2017}\natexlab{}.
\newblock \showarticletitle{Counterfactual Fairness}. In
  \bibinfo{booktitle}{{\em Advances in Neural Information Processing Systems
  30: Annual Conference on Neural Information Processing Systems 2017, 4-9
  December 2017, Long Beach, CA, {USA}}}. \bibinfo{pages}{4066--4076}.
\newblock


\bibitem[\protect\citeauthoryear{Lattimore and Szepesv{\'a}ri}{Lattimore and
  Szepesv{\'a}ri}{2020}]%
        {lattimore2020bandit}
\bibfield{author}{\bibinfo{person}{Tor Lattimore} {and} \bibinfo{person}{Csaba
  Szepesv{\'a}ri}.} \bibinfo{year}{2020}\natexlab{}.
\newblock \bibinfo{booktitle}{{\em Bandit algorithms}}.
\newblock \bibinfo{publisher}{Cambridge University Press}.
\newblock


\bibitem[\protect\citeauthoryear{Lee}{Lee}{2018}]%
        {Lee:2018ik}
\bibfield{author}{\bibinfo{person}{Min~Kyung Lee}.}
  \bibinfo{year}{2018}\natexlab{}.
\newblock \showarticletitle{{Understanding perception of algorithmic decisions:
  Fairness, trust, and emotion in response to algorithmic management}}.
\newblock \bibinfo{journal}{{\em Big Data {\&} Society\/}} \bibinfo{volume}{5},
  \bibinfo{number}{1} (\bibinfo{date}{March} \bibinfo{year}{2018}),
  \bibinfo{pages}{205395171875668--16}.
\newblock


\bibitem[\protect\citeauthoryear{MacKenzie}{MacKenzie}{2016}]%
        {MacKenzie:2016id}
\bibfield{author}{\bibinfo{person}{Donald~A. MacKenzie}.}
  \bibinfo{year}{2016}\natexlab{}.
\newblock \showarticletitle{{Performing Theory?}}
\newblock In \bibinfo{booktitle}{{\em An Engine, Not a Camera}}.
  \bibinfo{publisher}{The MIT Press}, \bibinfo{pages}{1--35}.
\newblock


\bibitem[\protect\citeauthoryear{Madaio, Stark, Wortman~Vaughan, and
  Wallach}{Madaio et~al\mbox{.}}{2020}]%
        {Madaio:2020cn}
\bibfield{author}{\bibinfo{person}{Michael~A Madaio}, \bibinfo{person}{Luke
  Stark}, \bibinfo{person}{Jennifer Wortman~Vaughan}, {and}
  \bibinfo{person}{Hanna Wallach}.} \bibinfo{year}{2020}\natexlab{}.
\newblock \showarticletitle{{Co-Designing Checklists to Understand
  Organizational Challenges and Opportunities around Fairness in AI}}. In
  \bibinfo{booktitle}{{\em CHI 2020: Proceedings of the SIGCHI Conference on
  Human Factors in Computing Systems}}. \bibinfo{address}{Honolulu, HI},
  \bibinfo{pages}{1--20}.
\newblock


\bibitem[\protect\citeauthoryear{Malik}{Malik}{2019}]%
        {Malik:2019ve}
\bibfield{author}{\bibinfo{person}{Momin~M. Malik}.}
  \bibinfo{year}{2019}\natexlab{}.
\newblock \bibinfo{title}{{Can algorithms themselves be biased?}}
\newblock   (\bibinfo{date}{April} \bibinfo{year}{2019}).
\newblock
\showURL{%
\url{https://medium.com/berkman-klein-center/can-algorithms-themselves-be-biased-cffecbf2302c}}


\bibitem[\protect\citeauthoryear{Malik}{Malik}{2020}]%
        {Malik:2020ws}
\bibfield{author}{\bibinfo{person}{Momin~M. Malik}.}
  \bibinfo{year}{2020}\natexlab{}.
\newblock \showarticletitle{{A Hierarchy of Limitations in Machine Learning}}.
\newblock \bibinfo{journal}{{\em arXiv\/}} (\bibinfo{date}{March}
  \bibinfo{year}{2020}), \bibinfo{pages}{1--68}.
\newblock


\bibitem[\protect\citeauthoryear{Mohamed, Png, and Isaac}{Mohamed
  et~al\mbox{.}}{2020}]%
        {Mohamed:2020eo}
\bibfield{author}{\bibinfo{person}{Shakir Mohamed},
  \bibinfo{person}{Marie-Therese Png}, {and} \bibinfo{person}{William Isaac}.}
  \bibinfo{year}{2020}\natexlab{}.
\newblock \showarticletitle{{Decolonial AI: Decolonial Theory as Sociotechnical
  Foresight in Artificial Intelligence}}.
\newblock \bibinfo{journal}{{\em Philosophy {\&} Technology\/}}
  \bibinfo{volume}{22}, \bibinfo{number}{4} (\bibinfo{date}{July}
  \bibinfo{year}{2020}), \bibinfo{pages}{16--28}.
\newblock


\bibitem[\protect\citeauthoryear{Narayanan}{Narayanan}{2018}]%
        {Narayanan:2018uz}
\bibfield{author}{\bibinfo{person}{Arvind Narayanan}.}
  \bibinfo{year}{2018}\natexlab{}.
\newblock \showarticletitle{{Translation tutorial: 21 fairness definitions and
  their politics}}. In \bibinfo{booktitle}{{\em FAT* 2018}}.
  \bibinfo{address}{New York}.
\newblock


\bibitem[\protect\citeauthoryear{Nissenbaum}{Nissenbaum}{2011}]%
        {Nissenbaum:2011uv}
\bibfield{author}{\bibinfo{person}{Helen Nissenbaum}.}
  \bibinfo{year}{2011}\natexlab{}.
\newblock \showarticletitle{{From Preemption to Circumvention}}.
\newblock \bibinfo{journal}{{\em Berkeley Technology Law Journal\/}}
  \bibinfo{volume}{26}, \bibinfo{number}{3} (\bibinfo{year}{2011}),
  \bibinfo{pages}{1367--1386}.
\newblock


\bibitem[\protect\citeauthoryear{Olteanu, Castillo, Diaz, and Kiciman}{Olteanu
  et~al\mbox{.}}{2019}]%
        {Olteanu:2019el}
\bibfield{author}{\bibinfo{person}{Alexandra Olteanu}, \bibinfo{person}{Carlos
  Castillo}, \bibinfo{person}{Fernando Diaz}, {and} \bibinfo{person}{Emre
  Kiciman}.} \bibinfo{year}{2019}\natexlab{}.
\newblock \showarticletitle{{Social Data: Biases, Methodological Pitfalls, and
  Ethical Boundaries}}.
\newblock \bibinfo{journal}{{\em Frontiers in Big Data\/}}  \bibinfo{volume}{2}
  (\bibinfo{date}{July} \bibinfo{year}{2019}), \bibinfo{pages}{1--33}.
\newblock


\bibitem[\protect\citeauthoryear{Pihkala and Karasti}{Pihkala and
  Karasti}{2016}]%
        {Pihkala:2016ca}
\bibfield{author}{\bibinfo{person}{Suvi Pihkala} {and} \bibinfo{person}{Helena
  Karasti}.} \bibinfo{year}{2016}\natexlab{}.
\newblock \showarticletitle{{Reflexive Engagement}}. In
  \bibinfo{booktitle}{{\em the 14th Participatory Design Conference}}.
  \bibinfo{publisher}{ACM Press}, \bibinfo{address}{New York, New York, USA},
  \bibinfo{pages}{21--30}.
\newblock


\bibitem[\protect\citeauthoryear{Potochnik}{Potochnik}{2012}]%
        {Potochnik:2012jb}
\bibfield{author}{\bibinfo{person}{Angela Potochnik}.}
  \bibinfo{year}{2012}\natexlab{}.
\newblock \showarticletitle{{Feminist implications of model-based science}}.
\newblock \bibinfo{journal}{{\em Studies in History and Philosophy of
  Science\/}} \bibinfo{volume}{43}, \bibinfo{number}{2} (\bibinfo{date}{June}
  \bibinfo{year}{2012}), \bibinfo{pages}{383--389}.
\newblock


\bibitem[\protect\citeauthoryear{Roemer}{Roemer}{2012}]%
        {Roemer:2012kj}
\bibfield{author}{\bibinfo{person}{John~E Roemer}.}
  \bibinfo{year}{2012}\natexlab{}.
\newblock \showarticletitle{{On Several Approaches to Equality of
  Opportunity}}.
\newblock \bibinfo{journal}{{\em Economics and Philosophy\/}}
  \bibinfo{volume}{28}, \bibinfo{number}{02} (\bibinfo{date}{Aug.}
  \bibinfo{year}{2012}), \bibinfo{pages}{165--200}.
\newblock


\bibitem[\protect\citeauthoryear{Rolland, Fitzgerald, Dings{\o}yr, and
  Stol}{Rolland et~al\mbox{.}}{2016}]%
        {RollandFDS16}
\bibfield{author}{\bibinfo{person}{Knut~H. Rolland}, \bibinfo{person}{Brian
  Fitzgerald}, \bibinfo{person}{Torgeir Dings{\o}yr}, {and}
  \bibinfo{person}{Klaas{-}Jan Stol}.} \bibinfo{year}{2016}\natexlab{}.
\newblock \showarticletitle{Problematizing Agile in the Large: Alternative
  Assumptions for Large-Scale Agile Development}. In \bibinfo{booktitle}{{\em
  Proceedings of the International Conference on Information Systems - Digital
  Innovation at the Crossroads, {ICIS} 2016, Dublin, Ireland, December 11-14,
  2016}}.
\newblock
\showURL{%
\url{http://aisel.aisnet.org/icis2016/ManagingIS/Presentations/7}}


\bibitem[\protect\citeauthoryear{Seaver}{Seaver}{2017}]%
        {Seaver:2017if}
\bibfield{author}{\bibinfo{person}{Nick Seaver}.}
  \bibinfo{year}{2017}\natexlab{}.
\newblock \showarticletitle{{Algorithms as culture: Some tactics for the
  ethnography of algorithmic systems}}.
\newblock \bibinfo{journal}{{\em Big Data {\&} Society\/}} \bibinfo{volume}{4},
  \bibinfo{number}{2} (\bibinfo{year}{2017}), \bibinfo{pages}{1--12}.
\newblock


\bibitem[\protect\citeauthoryear{Selbst, boyd, Friedler, Venkatasubramanian,
  and Vertesi}{Selbst et~al\mbox{.}}{2019}]%
        {Selbst:2019wf}
\bibfield{author}{\bibinfo{person}{Andrew~D. Selbst}, \bibinfo{person}{danah
  boyd}, \bibinfo{person}{Sorelle~A Friedler}, \bibinfo{person}{Suresh
  Venkatasubramanian}, {and} \bibinfo{person}{Janet Vertesi}.}
  \bibinfo{year}{2019}\natexlab{}.
\newblock \showarticletitle{{Fairness and Abstraction in Sociotechnical
  Systems}}. In \bibinfo{booktitle}{{\em Proceedings of the Conference on
  Fairness, Accountability, and Transparency}}. \bibinfo{publisher}{Association
  for Computing Machinery}, \bibinfo{address}{New York, NY, USA},
  \bibinfo{pages}{59--68}.
\newblock


\bibitem[\protect\citeauthoryear{Sengers, Boehner, David, and Kaye}{Sengers
  et~al\mbox{.}}{2005}]%
        {Sengers:2005uk}
\bibfield{author}{\bibinfo{person}{Phoebe Sengers}, \bibinfo{person}{Kirsten
  Boehner}, \bibinfo{person}{Shay David}, {and}
  \bibinfo{person}{Joseph~"Jofish" Kaye}.} \bibinfo{year}{2005}\natexlab{}.
\newblock \showarticletitle{{Reflective Design}}. In \bibinfo{booktitle}{{\em
  Proceedings of the th Decennial Aarhus Conference}}.
  \bibinfo{address}{Aarhus, Denmark}, \bibinfo{pages}{49--58}.
\newblock


\bibitem[\protect\citeauthoryear{Shilton}{Shilton}{2013}]%
        {Shilton:2013dd}
\bibfield{author}{\bibinfo{person}{Katie Shilton}.}
  \bibinfo{year}{2013}\natexlab{}.
\newblock \showarticletitle{{Values Levers: Building Ethics into Design}}.
\newblock \bibinfo{journal}{{\em Science, Technology, {\&} Human Values\/}}
  \bibinfo{volume}{38}, \bibinfo{number}{3} (\bibinfo{date}{May}
  \bibinfo{year}{2013}), \bibinfo{pages}{374--397}.
\newblock


\bibitem[\protect\citeauthoryear{Shilton}{Shilton}{2018a}]%
        {Shilton:2017hh}
\bibfield{author}{\bibinfo{person}{Katie Shilton}.}
  \bibinfo{year}{2018}\natexlab{a}.
\newblock \showarticletitle{{Engaging Values Despite Neutrality}}.
\newblock \bibinfo{journal}{{\em Science, Technology, {\&} Human Values\/}}
  \bibinfo{volume}{43}, \bibinfo{number}{2} (\bibinfo{year}{2018}),
  \bibinfo{pages}{247--269}.
\newblock


\bibitem[\protect\citeauthoryear{Shilton}{Shilton}{2018b}]%
        {Shilton:2018fb}
\bibfield{author}{\bibinfo{person}{Katie Shilton}.}
  \bibinfo{year}{2018}\natexlab{b}.
\newblock \showarticletitle{{Values and Ethics in Human-Computer Interaction}}.
\newblock \bibinfo{journal}{{\em Foundations and Trends{\textregistered} in
  Human{\textendash}Computer Interaction\/}} \bibinfo{volume}{12},
  \bibinfo{number}{2} (\bibinfo{year}{2018}), \bibinfo{pages}{107--171}.
\newblock


\bibitem[\protect\citeauthoryear{Simonsen and Robertson}{Simonsen and
  Robertson}{2013}]%
        {Simonsen:2013ws}
\bibfield{editor}{\bibinfo{person}{Jesper Simonsen} {and} \bibinfo{person}{Toni
  Robertson}} (Eds.). \bibinfo{year}{2013}\natexlab{}.
\newblock \bibinfo{booktitle}{{\em {Routledge International Handbook of
  Participatory Design}}}.
\newblock \bibinfo{publisher}{Routledge}, \bibinfo{address}{New York and
  London}.
\newblock


\bibitem[\protect\citeauthoryear{Sloan, Moss, Awomolo, and Forlano}{Sloan
  et~al\mbox{.}}{2020}]%
        {Sloan:2020um}
\bibfield{author}{\bibinfo{person}{Mona Sloan}, \bibinfo{person}{Emanuel Moss},
  \bibinfo{person}{Olaitan Awomolo}, {and} \bibinfo{person}{Laura Forlano}.}
  \bibinfo{year}{2020}\natexlab{}.
\newblock \showarticletitle{{Participation is not a Design Fix for Machine
  Learning}}. In \bibinfo{booktitle}{{\em Proceedings of the the International
  Conference on Machine Learning}}. \bibinfo{address}{Vienna, Austria},
  \bibinfo{pages}{1--7}.
\newblock


\bibitem[\protect\citeauthoryear{Stark}{Stark}{2019}]%
        {Stark:2019fs}
\bibfield{author}{\bibinfo{person}{Luke Stark}.}
  \bibinfo{year}{2019}\natexlab{}.
\newblock \showarticletitle{{Facial recognition is the plutonium of AI}}.
\newblock \bibinfo{journal}{{\em XRDS: Crossroads, The ACM Magazine for
  Students\/}} \bibinfo{volume}{25}, \bibinfo{number}{3} (\bibinfo{date}{April}
  \bibinfo{year}{2019}), \bibinfo{pages}{50--55}.
\newblock


\bibitem[\protect\citeauthoryear{Stark and Tierney}{Stark and Tierney}{2014}]%
        {Stark:2013bl}
\bibfield{author}{\bibinfo{person}{Luke Stark} {and} \bibinfo{person}{Matt
  Tierney}.} \bibinfo{year}{2014}\natexlab{}.
\newblock \showarticletitle{{Lockbox: mobility, privacy and values in cloud
  storage}}.
\newblock \bibinfo{journal}{{\em Ethics and Information Technology\/}}
  \bibinfo{volume}{16}, \bibinfo{number}{1} (\bibinfo{date}{March}
  \bibinfo{year}{2014}), \bibinfo{pages}{1--13}.
\newblock


\bibitem[\protect\citeauthoryear{Suchman}{Suchman}{2006}]%
        {Suchman:2006uq}
\bibfield{author}{\bibinfo{person}{Lucy Suchman}.}
  \bibinfo{year}{2006}\natexlab{}.
\newblock \bibinfo{booktitle}{{\em {Human-Machine Reconfigurations: Plans and
  Situated Actions}}}.
\newblock \bibinfo{publisher}{Cambridge University Press},
  \bibinfo{address}{Cambridge, UK}.
\newblock


\bibitem[\protect\citeauthoryear{Suchman}{Suchman}{2002}]%
        {Suchman:2002tp}
\bibfield{author}{\bibinfo{person}{Lucy~A Suchman}.}
  \bibinfo{year}{2002}\natexlab{}.
\newblock \showarticletitle{{Located Accountabilities in Technology
  Production.}}
\newblock \bibinfo{journal}{{\em Scand. J. Inf. Syst.\/}}
  (\bibinfo{year}{2002}).
\newblock


\bibitem[\protect\citeauthoryear{Verma and Rubin}{Verma and Rubin}{2018}]%
        {Verma:2018hw}
\bibfield{author}{\bibinfo{person}{Sahil Verma} {and} \bibinfo{person}{Julia
  Rubin}.} \bibinfo{year}{2018}\natexlab{}.
\newblock \showarticletitle{{Fairness definitions explained}}. In
  \bibinfo{booktitle}{{\em 2018 ACM/IEEE International Workshop on Software
  Fairness}}. \bibinfo{publisher}{ACM Press}, \bibinfo{address}{New York, New
  York, USA}, \bibinfo{pages}{1--7}.
\newblock


\bibitem[\protect\citeauthoryear{Wang, Zhang, Liu, Liu, and Miao}{Wang
  et~al\mbox{.}}{2018}]%
        {Wang:2018ht}
\bibfield{author}{\bibinfo{person}{Cunrui Wang}, \bibinfo{person}{Qingling
  Zhang}, \bibinfo{person}{Wanquan Liu}, \bibinfo{person}{Yu Liu}, {and}
  \bibinfo{person}{Lixin Miao}.} \bibinfo{year}{2018}\natexlab{}.
\newblock \showarticletitle{{Facial feature discovery for ethnicity
  recognition}}.
\newblock \bibinfo{journal}{{\em WIREs Data Mining and Knowledge Discovery\/}}
  \bibinfo{volume}{9}, \bibinfo{number}{1} (\bibinfo{date}{Aug.}
  \bibinfo{year}{2018}), \bibinfo{pages}{85--17}.
\newblock


\bibitem[\protect\citeauthoryear{Woodruff, Fox, Rousso-Schindler, and
  Warshaw}{Woodruff et~al\mbox{.}}{2018}]%
        {Woodruff:2018if}
\bibfield{author}{\bibinfo{person}{Allison Woodruff}, \bibinfo{person}{Sarah~E
  Fox}, \bibinfo{person}{Steven Rousso-Schindler}, {and}
  \bibinfo{person}{Jeffrey Warshaw}.} \bibinfo{year}{2018}\natexlab{}.
\newblock \showarticletitle{{A Qualitative Exploration of Perceptions of
  Algorithmic Fairness}}. In \bibinfo{booktitle}{{\em Extended Abstracts of the
  2018 CHI Conference}}. \bibinfo{publisher}{ACM Press}, \bibinfo{address}{New
  York, New York, USA}, \bibinfo{pages}{1--14}.
\newblock


\end{thebibliography}

\end{document}